\documentclass[a4paper,11pt]{article}
\pdfoutput=1 

\usepackage{jheppub} 

\usepackage[T1]{fontenc}
\usepackage{multirow}
\usepackage{pagecolor}

\allowdisplaybreaks
 
\definecolor{nicered}{rgb}{0.7,0.1,0.1}
\definecolor{nicegreen}{rgb}{0.1,0.5,0.1}
\definecolor{niceblue}{rgb}{0.0,0.1,0.7}
\hypersetup{colorlinks,citecolor=niceblue,linkcolor=niceblue,urlcolor=niceblue}

\usepackage[normalem]{ulem}

\def \bm#1{\mbox{\boldmath$#1$\unboldmath}}
\def \beq{\begin{equation}}
\def \eeq{\end{equation}}
\def \bea{\begin{eqnarray}}
\def \eea{\end{eqnarray}}

\preprint{MPP-2022-74}

\title{\boldmath Drell-Yan production in third-generation gauge vector leptoquark models at NLO$+$PS in QCD}

\author[a]{Ulrich Haisch,}
\author[a,\, b]{Luc Schnell}
\author[a,\, b]{and Stefan Schulte}

\affiliation[a]{Max Planck Institute for Physics, \\ F{\"o}hringer Ring 6, 80805 M{\"u}nchen, Germany}
\affiliation[b]{Technische Universit{\"a}t M{\"u}nchen, Physik-Department, \\ James-Franck-Strasse 1, 85748 Garching, Germany}

\emailAdd{haisch@mpp.mpg.de}
\emailAdd{schnell@mpp.mpg.de}
\emailAdd{sschulte@mpp.mpg.de}

\abstract{Motivated by the long-standing hints of lepton-flavour non-universality in the~$b \to c \ell \nu$ and~$b \to s \ell^+ \ell^-$ channels, we study Drell-Yan ditau production at the Large~Hadron~Collider~(LHC). In the context of models with third-generation gauge vector leptoquarks~(LQs), we calculate the complete~${\cal O} (\alpha_s)$ corrections to the~$pp \to \tau^+ \tau^-$ process, achieving next-to-leading order~(NLO) plus parton~shower~(NLO$+$PS) accuracy using the {\tt POWHEG} method. We~provide a dedicated Monte~Carlo code that evaluates the NLO QCD corrections on-the-fly in the event generation and use it to study the numerical impact of NLO$+$PS corrections on the kinematic distributions that enter the existing experimental searches for non-resonant ditau final states. Based on our phenomenological analysis we derive NLO accurate constraints on the masses and couplings of third-generation gauge vector~LQs~using the latest LHC ditau search results corresponding to an integrated luminosity of around~$140 \, {\rm fb}^{-1}$ of proton-proton collisions at~$\sqrt{s} = 13 \, {\rm TeV}$. The presented~NLO$+$PS generator allows for an improved signal modelling, making it an essential tool for future ATLAS and CMS searches for vector~LQs~in~$\tau^+ \tau^-$ final states at LHC~Run~III and~beyond.}

\begin{document} 
\pagecolor{white}
\maketitle
\flushbottom

\section{Introduction}
\label{sec:introduction}

The prevailing hints of lepton-flavour universality~(LFU) violation that have been observed in both the~$b \to c\ell \nu$~\cite{BaBar:2012obs,BaBar:2013mob,LHCb:2015gmp,LHCb:2017smo,LHCb:2017rln,Belle:2019gij} and~$b \to s \ell^+ \ell^-$~\cite{LHCb:2017avl,LHCb:2019hip,Belle:2019oag,BELLE:2019xld,LHCb:2021trn} transitions are commonly considered the most compelling departures from the Standard Model~(SM) observed by collider experiments in recent years. Thanks to a concerted theoretical effort~\cite{Alonso:2015sja,Calibbi:2015kma,Fajfer:2015ycq,Barbieri:2015yvd,Hiller:2016kry,Bhattacharya:2016mcc,Barbieri:2016las,Buttazzo:2017ixm,Assad:2017iib,DiLuzio:2017vat,Calibbi:2017qbu,Bordone:2017bld,Barbieri:2017tuq,Blanke:2018sro,Greljo:2018tuh,Bordone:2018nbg,Kumar:2018kmr,Azatov:2018kzb,DiLuzio:2018zxy,Angelescu:2018tyl,Schmaltz:2018nls,Fornal:2018dqn,Aebischer:2019mlg,Cornella:2019hct,Shi:2019gxi,DaRold:2019fiw,Bordone:2019uzc,Crivellin:2019dwb,Altmannshofer:2020ywf,Fuentes-Martin:2020bnh,Guadagnoli:2020tlx,Iguro:2020keo,Alda:2020okk,Bhaskar:2021pml,Iguro:2021kdw,Angelescu:2021lln,Cornella:2021sby,Belfatto:2021ats,Barbieri:2022ikw} it has been established that singlet vector leptoquarks (LQs) with a mass in the TeV range and third-generation couplings provide a simple, especially appealing explanation of these flavour~anomalies. 

Several different search strategies for third-generation~LQs~have so far been considered at the Large Hadron Collider~(LHC). While the ATLAS and CMS collaborations have initially focused on strong~LQ~pair production in gluon-gluon fusion or quark-antiquark annihilation, recently also single~LQ~production in gluon-quark fusion and~$t$-channel~LQ~exchange in Drell-Yan~(DY)~dilepton production have been exploited to constrain the~LQ-quark-lepton couplings. See~\cite{ATLAS:2020zms,ATLAS:2020dsk,CMS:2020wzx,CMS:2022goy,CMS-PAS-EXO-19-016} for the latest experimental results of these~kinds. Resonant~LQ~signatures arising from quark-lepton annihilation at the~LHC~\cite{Buonocore:2020erb,Buonocore:2020nai,Greljo:2020tgv,Buonocore:2022msy} have also been studied and found to provide complementary information compared to the other third-generation~LQ~search~strategies~\cite{Haisch:2020xjd}. 

In the context of the singlet vector~LQ~model,~LHC~searches for non-resonant ditau final states have been shown to be particularly important~\cite{Faroughy:2016osc,Schmaltz:2018nls,Baker:2019sli,Angelescu:2021lln,Bhaskar:2021pml,Cornella:2021sby,CMS:2022goy}.\footnote{Further detailed investigations of other non-resonant phenomena in~DY~production related to the semi-leptonic~$B$-decay anomalies can be found in the articles~\cite{Raj:2016aky,Greljo:2017vvb,Allanach:2017bta,Dorsner:2018ynv,Afik:2018nlr,Bansal:2018eha,Allanach:2018odd,Mandal:2018kau,Choudhury:2019ucz,Angelescu:2020uug,Crivellin:2021rbf,Crivellin:2021egp,Crivellin:2021bkd,Garland:2021ghw,Crivellin:2022mff,Azatov:2022itm,Allwicher:2022gkm}.} Given~the relevance of the~$pp \to \tau^+ \tau^-$ process, the main goal of this article is to improve the theoretical description of~DY~dilepton production in models with a singlet vector~LQ~by calculating the relevant next-to-leading order~(NLO) corrections in QCD. These~fixed-order predictions are then consistently matched to a parton shower~(PS) utilising the~{\tt POWHEG}~method~\cite{Nason:2004rx,Frixione:2007vw} as implemented in the {\tt POWHEG-BOX}~\cite{Alioli:2010xd}. This~allows for a realistic exclusive description of~DY~dilepton~processes in singlet vector~LQ~models at the level of hadronic events. Similar~calculations have been performed in the case of scalar~LQs~in~\cite{Alves:2018krf,Haisch:2022lkt} and the work presented in the following constitutes a non-trivial extension of our previous article~\cite{Haisch:2022lkt}. The~added complications that arise here are related to the fact that unambiguous~NLO~QCD calculations are only possible in the case of a massive vector~LQ~if the corresponding~field is embedded into a consistent ultraviolet~(UV)~complete model. An~inescapable consequence of such an embedding is the presence of additional states like for example colorons that carry non-zero~$SU(3)_C$ charges and have masses close to that of the vector~LQ~\cite{DiLuzio:2017vat,Baker:2019sli}. As~stressed~in~the~second part of the trilogy~\cite{Fuentes-Martin:2019ign,Fuentes-Martin:2020luw,Fuentes-Martin:2020hvc}, a proper treatment of all~${\cal O} (\alpha_s)$~corrections is therefore necessary to determine the full~NLO~QCD contributions and that calculations such as~\cite{Aebischer:2018acj} that include only the corrections associated to virtual and real~QCD~emissions~may lead to inaccurate~results in realistic third-generation~vector~LQ models. In order to obtain the proper~${\cal O} (\alpha_s)$~corrections~to~DY~dilepton production in vector~LQ~models our~NLO$+$PS~{\tt POWHEG-BOX} implementation therefore contains the contributions from virtual and real gluons as well as coloron loops. The obtained analytic expressions furthermore serve as an independent cross check of the computations presented in the publication~\cite{Fuentes-Martin:2020luw}. 

In our phenomenological analysis we discuss the numerical impact of the NLO~QCD corrections on the kinematic distributions that enter the existing ATLAS and CMS searches for non-resonant phenomena in~$\tau^+ \tau^-$ final states. Since it is known that the requirement of additional final-state jets containing the decay of a~$B$~hadron~($b\hspace{0.4mm}$-jets) helps to improve the~LHC~sensitivity of third-generation~LQ~searches~\cite{ATLAS:2020zms,ATLAS:2021mla,CMS:2022goy,Afik:2018nlr,Choudhury:2019ucz,Altmannshofer:2017poe,Iguro:2017ysu,Abdullah:2018ets,Marzocca:2020ueu,Endo:2021lhi,Haisch:2022lkt,CMS-PAS-EXO-19-016} we pay special attention to this feature in our study. Based on our~DY~ditau~analyses we are able to derive improved limits on the parameter space of third-generation singlet vector~LQ models using the results~\cite{CMS:2022goy} that utilise the full~LHC~Run~II integrated luminosity of~$138 \, {\rm fb}^{-1}$ obtained for proton-proton~($pp$) collisions at a centre-of-mass~energy of~$\sqrt{s} = 13 \, {\rm TeV}$. We~also consider the constraints on the parameter space of third-generation singlet vector~LQs that are imposed by the recent LHC~Run~II searches~\cite{ATLAS:2020zms,CMS-PAS-EXO-19-016} for ditau~production in our supplementary material. 

This article is organised as follows. In Section~\ref{sec:theory} we specify the structure of the~vector~LQ~interactions that we consider in this work. Section~\ref{sec:calculation}~briefly describes the basic ingredients of the NLO~QCD calculation of the DY~dilepton process and their implementation into the {\tt POWHEG-BOX}. The impact of the NLO$+$PS corrections on the kinematic distributions in~$pp \to \tau^+ \tau^-$ production is presented in~Section~\ref{sec:analysis}. Our recast of the~LHC~search~\cite{CMS:2022goy} is discussed in~Section~\ref{sec:limits}, where we also derive improved limits on the~LQ-quark-lepton couplings and masses of third-generation singlet vector~LQs. Section~\ref{sec:conclusions} contains our conclusions. Additional material is relegated to three appendices. In Appendix~\ref{app:feynman}~we spell out the form of the pure gauge, Goldstone boson and ghost interactions needed to perform the calculation of the third-generation gauge vector~LQ corrections considered in this work. The constraints on the parameter space of third-generation singlet vector~LQs that follow from recasts of the recent ditau searches~\cite{ATLAS:2020zms,CMS-PAS-EXO-19-016} are instead presented in~Appendix~\ref{app:moreconstraints}. For~the sake of completeness, Appendix~\ref{app:Zprime} contains a brief study of the impact of $Z^\prime$~exchange in DY ditau production. So~without further ado, let's crack straight into it.

\section{Theoretical framework}
\label{sec:theory}

A singlet vector~LQ~can be added to the~SM Lagrangian in a simple bottom-up approach by employing the following effective interactions
\begin{equation} \label{eq:U1_fermion_interactions}
\mathcal{L}_{U} \supset \, \frac{g_U}{\sqrt{2}} \left[\beta_L^{ij} \, \bar{Q}^{\hspace{0.4mm} i, a} \hspace{0.4mm} \gamma_\mu \hspace{0.4mm} L^j + \beta_R^{ij} \, \bar{d}^{\hspace{0.4mm} i, a} \hspace{0.4mm} \gamma_\mu\hspace{0.4mm} e^{\hspace{0.4mm} j} \right] U^{\mu, a} + {\rm h.c.}
\end{equation}
Here~$Q$~and~$L$ are the left-handed~SM~quark and lepton $SU(2)_L$~doublets, while~$d$ and~$e$ are the corresponding right-handed~fields,~$i,j \in \{1, 2 ,3\}$ are flavour indices and~$a \in \{1, 2, 3\}$ is a colour index. The vector~LQ~transforms as~$U \sim \left(3, 1, 2/3\right)$ under~the~SM~gauge group~$G_{\rm SM} = SU(3)_C \times SU(2)_L \times U(1)_Y$, making it an~$SU(2)_L$ singlet. The coupling~$g_U$~characterises the overall strength of the~LQ~interactions with the~SM matter fields, whereas~$\beta^{ij}_{L}$ and~$\beta^{ij}_{R}$ are (a~priori) arbitrary complex~$3\times 3$ matrices in flavour space.\footnote{In our {\tt POWHEG-BOX} implementation of~the simplified Lagrangian~(\ref{eq:U1_fermion_interactions}) the relevant third-generation LQ-quark-lepton couplings are treated as real.}  In order to explain the observed anomalies in the charged-current~$b \to c$ and neutral-current~$b \to s$ transitions the~following LQ-quark-lepton~couplings have to be non-zero and follow the pattern~$\left |\beta_L^{33} \right | \simeq \left |\beta_R^{33} \right | \gtrsim \left |\beta_L^{23} \right | \gg \left |\beta_L^{32} \right | \simeq \left |\beta_L^{22} \right |$, while the remaining couplings can in principle~vanish. 

The simplified interactions~described by the Lagrangian~${\cal L}_{U}$~however do not provide a consistent~UV~completion for the singlet vector~LQ~field which renders higher-order perturbative calculations based on~(\ref{eq:U1_fermion_interactions}) in general ambiguous. A well-motivated and thoroughly studied class of UV-complete theories that incorporates a singlet vector~LQ~are gauge models. There, the massive~$U$ field arises from a gauge symmetry~$G \supset G_\text{SM}$ that is broken spontaneously to yield the~SM Lagrangian at low energies, together with the singlet vector~LQ~as well as additional degrees of freedom. The minimal gauge group that leads to the effective interactions of the form~(\ref{eq:U1_fermion_interactions}) and that can account for the hints of~LFU violation in semi-leptonic~$B$ decays~is~\cite{DiLuzio:2017vat,Bordone:2017bld,DiLuzio:2018zxy,Bordone:2018nbg,Greljo:2018tuh,Cornella:2019hct,Fuentes-Martin:2020bnh,Guadagnoli:2020tlx,Georgi:2016xhm,Diaz:2017lit}
\begin{equation} \label{eq:G4321}
G_\text{4321} = SU(4) \times SU(3)^\prime \times SU(2)_L \times U(1)_{X}\,. 
\end{equation}
This gauge group is commonly referred to as 4321. In our article, we restrict ourselves to the~$SU(4) \times SU(3)^\prime$ sector of~(\ref{eq:G4321}) which includes the~LQ~interactions and~$\mathcal{O}(\alpha_s)$ corrections thereof, while neglecting contributions that involve the~$SU(2)_L \times U(1)_X$ subgroup. This~means in particular that we do not consider contributions to DY dilepton production that arise from the colour singlet state~$Z^\prime \sim \left(1,1, 0 \right)$ that also appears in the spectrum of the 4321 model after spontaneous symmetry breaking~\cite{DiLuzio:2017vat,Bordone:2017bld,DiLuzio:2018zxy,Bordone:2018nbg,Greljo:2018tuh,Cornella:2019hct}. This omission is firstly motivated because the~$Z^\prime$ does not contribute to~the~${\cal O} (\alpha_s)$ corrections we are interested in. Secondly, while the colour singlet does contribute to DY~dilepton~production, the tree-level~$s$-channel exchange of a~$Z^\prime$ leads to a narrow resonance in the dilepton invariant mass spectrum of~$pp \to \ell^+ \ell^-$. In contrast, the leading contribution to DY~dilepton~production due to~(\ref{eq:U1_fermion_interactions}) corresponds to a non-resonant signal associated to~$t$-channel exchange of the singlet vector LQ. Since experimentally resonant DY~dilepton~signatures can in principle be disentangled from non-resonant ones, treating the~$Z^\prime$ and the~$U$ contributions also separately in a theoretical analysis seems justified. In Appendix~\ref{app:Zprime} we dwell further on this point, considering the ditau final state as an example.

In the 4321 model, the symmetry~(\ref{eq:G4321}) is broken spontaneously via two scalars once these fields acquire non-zero vacuum expectation values. The massive~$U$ field results from the broken~$SU(4)$ group alone, while the~$SU(4)$ and~$SU(3)^\prime$ groups conspire to yield the~SM~gluon~$G$ and an additional massive colour-octet vector~$G^{\hspace{0.2mm} \prime} \sim \left( 8, 1, 0 \right)$, commonly referred to as coloron. Explicitly, one has in the case of the singlet vector LQ 
\begin{equation} \label{eq:LQ4321}
U_{\mu}^{1,2,3} = \frac{1}{\sqrt{2}} \, \Big( H_\mu^{\hspace{0.2mm} 9,11,13} - i \hspace{0.2mm} H_\mu^{\hspace{0.2mm} 10, 12, 14} \Big) \,, \\[2mm]
\end{equation}
where~$H_\mu^A$ with~$A \in \left \{ 1, \dots, 15 \right \}$ are the~$SU(4)$ gauge fields. The colour octet states,~i.e.~the SM gluon and the coloron, are instead given by the following linear combinations 
\begin{equation} \label{eq:octets4321}
G_\mu^a = s_3 \hspace{0.4mm} H_\mu^a+ c_3 \hspace{0.4mm} C_\mu^a \, , \qquad 
G_\mu^{\hspace{0.2mm} \prime \hspace{0.2mm} a} = c_3 \hspace{0.4mm} H_\mu^a - s_3 \hspace{0.4mm} C_\mu^a \,, 
\end{equation}
with~$C_\mu^a$ the~$SU(3)^\prime$~gauge fields and we have introduced the following abbreviations 
\begin{equation} \label{eq:s3c3}
s_3 = \sin \theta_3 = \frac{g_3}{\sqrt{g_4^2 + g_3^2}} \,, \qquad 
c_3 = \cos \theta_3 = \frac{g_4}{\sqrt{g_4^2 + g_3^2}} \,, 
\end{equation}
for the sine and cosine of the mixing angle~$\theta_3$ in the~$SU(4) \times SU(3)^\prime$ sector. Here~$g_4$~($g_3$) denotes the coupling constant associated to the~$SU(4)$~$\big( SU(3)^\prime \big)$ group. The strong QCD coupling constant~$g_s$ can be expressed in terms of~$g_4$,~$g_3$ and (\ref{eq:s3c3}) as 
\begin{equation} \label{eq:gs}
g_s = s_3 \hspace{0.4mm} g_4 = c_3 \hspace{0.4mm} g_3 \,. 
\end{equation}

The large couplings of the singlet vector LQ to the third quark family, as required to explain the~$B$-decay anomalies, is achieved by unifying the third fermion generation~(and~only the third) into~$SU(4)$ quadruplets. Specifically, the SM fermion fields then take the form~$\Psi_L= \left(Q^3, L^3 \right)^T$ and~$\Psi_R^- = \left( d^{\hspace{0.2mm} 3}, e^3 \right)^T$, which we will from now on generically denote by~$\Psi = \left( \psi_q, \psi_\ell \right)^T$. This representation transforms as~$\Psi \sim (4, 1)$ under the~$SU(4) \times SU(3)^\prime$ gauge~group. After spontaneous symmetry breaking, the interactions between the coloured gauge bosons and the third-generation fermions in the 4321 model then read
\begin{equation} \label{eq:4321_gauge_fermion}
\begin{split}
\mathcal{L}_{4321} & \supset \, \frac{g_4}{\sqrt{2}} \, \bar{\psi}_q^a \hspace{0.4mm} \gamma_\mu \hspace{0.4mm} \psi_\ell \hspace{0.75mm} U^{\mu, a} + {\rm h.c.} 
+ g_s \hspace{0.4mm} \bar{\psi}_q \hspace{0.4mm} \gamma_\mu T^a \hspace{0.4mm} \psi_q \hspace{0.5mm} G^{\hspace{0.2mm} \mu, a} + c_3 \hspace{0.4mm} g_4\, \bar{\psi}_q \hspace{0.4mm} \gamma_\mu T^a \hspace{0.4mm} \psi_q \hspace{0.5mm} G^{\hspace{0.2mm} \prime \hspace{0.2mm} \mu, \hspace{0.2mm} a} \,,
\end{split}
\end{equation}
where~the symbol $T^a$ denotes the usual~$SU(3)$ generators. Notice that the first two terms in~(\ref{eq:4321_gauge_fermion}) resemble the effective singlet vector LQ interactions~(\ref{eq:U1_fermion_interactions}) if one identifies~$g_U = g_4$ and~$\beta^{33}_L = \beta^{33}_R = 1$, which shows that ${\cal L}_{U}$ is correctly recovered if the $U$ field is embedded into the~4321~model. As~a result of the enlarged gauge group the 4321 model however contains besides a massless gluon $G$ also a massive coloron $G^{\hspace{0.2mm} \prime}$ that couples to the SM third-generation quarks with strength~$c_3 \hspace{0.4mm} g_4$. This implies that one-loop amplitudes in the full~4321~theory in general receive ${\cal O} (\alpha_s)$ contributions from both virtual $G$ and $G^\prime$ exchange. In fact, for any given process the gluon-mediated amplitude is proportional to $g_s^2$, while the coloron-mediated amplitude is proportional to $\left ( c_3 \hspace{0.4mm} g_4 \right )^2 = g_4^2 - g_s^2$. Notice that the minus sign in this relation ensures a perfect cancellation of UV~divergences proportional to $g_s^2$. This shows that in the 4321 model coloron effects necessarily have to be included if one wants to correctly calculate scattering processes such as $b \bar b \to \tau^+ \tau^-$~beyond the leading order~(LO) in QCD. 

\section{Calculation in a nutshell}
\label{sec:calculation}

\begin{figure}[t!]
\begin{center}
\includegraphics[width=0.75\textwidth]{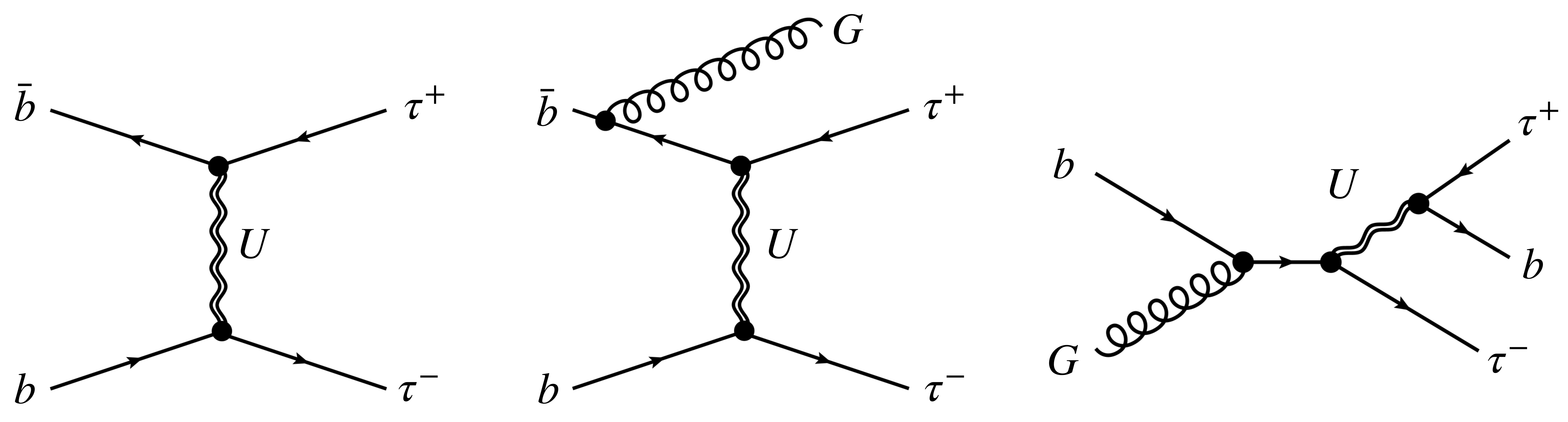}
\end{center}
\vspace{-2mm} 
\caption{\label{fig:diagrams1} Examples of~singlet vector~LQ contributions to the~DY~ditau spectrum initiated by bottom-quark fusion. The left Feynman diagram describes the tree-level process involving~$t$-channel~singlet vector~LQ exchange~($U$), while the middle~(right) graph represents the real gluon~($G$) corrections with non-resonant~(resonant) intermediate~$U$. See main text for further details. }
\end{figure}

\begin{figure}[t!]
\vspace{6mm} 
\begin{center}
\includegraphics[width=0.95\textwidth]{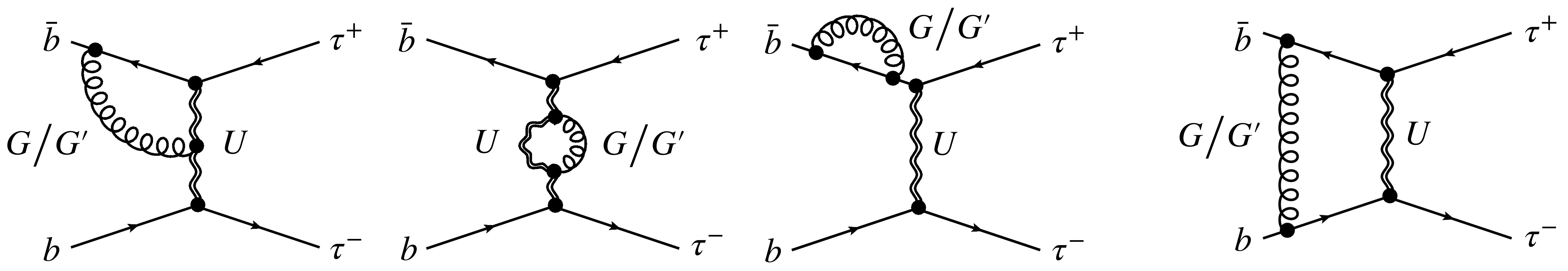}
\end{center}
\vspace{-2mm} 
\caption{\label{fig:diagrams2} Virtual~$\mathcal{O}(\alpha_s)$ corrections to the~singlet vector~LQ~contribution in~DY~ditau production, with a gluon~($G$) or a coloron~($G^{\hspace{0.2mm} \prime}$) running in the loop. The three graphs on the left show the factorisable contributions. They arise from LQ-quark-lepton vertex corrections as well as from LQ and quark wave function corrections. The diagram on the far right depicts a non-factorisable contribution due to a box diagram. For additional explanations consult the main text.}
\end{figure}

Representative Feynman diagrams leading to~DY~ditau production in the presence of~(\ref{eq:4321_gauge_fermion}) are displayed in Figures~\ref{fig:diagrams1} and~\ref{fig:diagrams2}. The first figure shows the tree-level process involving~$t$-channel~singlet vector~LQ~exchange~(left) and the corresponding real gluon~corrections~(middle and right). Notice that all depicted contributions are initiated by bottom-quark~($b \bar b$) fusion.\footnote{Throughout this article we work in the five-flavour scheme, where charm- and bottom-quarks are considered as partons in the proton and as such have a corresponding parton distribution function~(PDF).} We include real contributions with both non-resonant~(middle) and resonant~(right) intermediate~$U$ states, the latter case corresponding to single-LQ~production with a subsequent decay of the~singlet vector~LQ~to a pair of a bottom quark and an anti-tau, i.e.~$Gb \to U \tau^-$ followed by~$U \to b \tau^+$. These resonant diagrams also contribute at~$\mathcal{O}(\alpha_s)$ and are particularly important for invariant ditau masses~($m_{\tau \tau}$) close to the singlet vector~LQ~mass~$M_{U}$. At the same time, we neglect~$\mathcal{O}(\alpha_s)$ corrections associated to real coloron emissions. This is theoretically justified because these contributions are, unlike the real gluon emissions, infrared~(IR) finite by themselves. Furthermore, the stringent bounds on the coloron mass~from~LHC searches for dijet and ditop production~\cite{Cornella:2021sby} that impose~$M_{G^{\hspace{0.2mm} \prime}} \gtrsim 3 \, {\rm TeV}$ are expected to render the resonant~$G^{\hspace{0.2mm} \prime}$ contribution to the $b \bar b \to \tau^+ \tau^-$ process insignificant for all practical purposes. 

In Figure~\ref{fig:diagrams2} we display an assortment of the virtual ${\cal O} (\alpha_s)$ contributions that are included in our calculation. The three factorisable corrections shown on the left exhibit UV~divergences, which only cancel if both the gluon and coloron contributions are included. This shows that the coloron contributions are intimately tied to the gluon corrections in the~4321~model. Notice that besides the interaction terms between the SM fermions and the coloured gauge bosons~(\ref{eq:4321_gauge_fermion}) also factorisable diagrams with vertices involving only coloured gauge bosons and graphs with Goldstone bosons and ghosts need to be considered if the computation is performed in the Feynman or any other renormalisable or $R_\xi$~gauge~(cf.~Appendix~\ref{app:feynman} for details). Last but not least, the process $b \bar b \to \tau^+ \tau^-$ receives finite contributions from the non-factorisable box diagram shown on the very right in Figure~\ref{fig:diagrams2}. 

Besides QCD corrections to the $b \bar b \to \tau^+ \tau^-$ process we also study in our article~the potential size of interference effects between the SM background and the~singlet vector LQ~signal. We treat these effects at the LO in perturbation theory, which means that our {\tt POWHEG-BOX} implementation of DY~dilepton~production contains the squared matrix elements built from the~SM~corrections involving~$Z$-boson or photon exchange in the~$s$-channel and the~$t$-channel~singlet~vector~LQ exchange contribution (cf.~the~left diagram in Figure~\ref{fig:diagrams1}).

In the calculation of the squared matrix elements, we use conventional dimensional regularisation for both~UV and IR singularities. For the generation and computation of the squared matrix elements, we rely on the {\tt Mathematica} packages {\tt FeynRules}~\cite{Alloul:2013bka}, {\tt FeynArts}~\cite{Hahn:2000kx}, {\tt FormCalc}~\cite{Hahn:2016ebn} and {\tt Package-X}~\cite{Patel:2015tea}, while making use of {\tt LoopTools}~\cite{Hahn:1998yk} for the numerical evaluation of the Passarino-Veltman integrals that appear in the one-loop contributions. Throughout this article we work in the on-shell scheme. To deal with the soft and collinear singularities of the real corrections to the~$t$-channel~singlet vector LQ~exchange contribution, cf.~the~middle diagram in~Figure~\ref{fig:diagrams1}, and to cancel the IR~poles of the one-loop virtual corrections, cf.~the~first and the third diagram in~Figure~\ref{fig:diagrams2}, we~exploit the Frixione-Kunszt-Signer subtraction~\cite{Frixione:1995ms,Frixione:1997np}. Specifically, we use the {\tt POWHEG-BOX} to automatically build the soft and collinear counterterms and remnants, also checking the behaviour of the real squared matrix elements in the soft and collinear limits against their soft and collinear approximations. Notice that the real NLO~QCD contributions that describe resonant single~production of a $U$ and its subsequent decay,~cf.~the~right diagram in~Figure~\ref{fig:diagrams1}, are IR~finite and hence do not require an IR subtraction. Our MC code therefore allows to achieve NLO+PS accuracy for~DY~dilepton production in~singlet vector LQ models. The~presented NLO$+$PS generator is in particular able to generate events with one additional QCD parton from the matrix element calculation without the need to introduce a spurious merging or matching scale. Two-jet events are instead exclusively generated by the PS in our MC setup. 

\section{Numerical applications}
\label{sec:analysis}

As a first application we calculate the $\mathcal{O}(\alpha_s)$ corrections to the~partial decay widths of the singlet vector LQ. Since a detailed description of this computation has already been given in the publication~\cite{Fuentes-Martin:2020luw} we do not repeat it here. Employing~(\ref{eq:4321_gauge_fermion}) and the pure gauge, Goldstone boson and ghost terms given~(\ref{eq:4321_other}),\footnote{Throughout our work we neglect the impact of radial modes. In the case of the partial decay widths this has been shown in~\cite{Fuentes-Martin:2020luw} to be an excellent numerical approximation in the limit $M_U, M_{G^\prime} \ll M_R$ with~$M_R$ denoting the common mass of the radial modes. }  we find for the process $U \to b \tau$ the analytic result
\begin{equation} \label{eq:GammaUbtau}
\Gamma \left ( U \to b \tau \right ) = \frac{g_4^2 M_U}{24 \hspace{0.125mm} \pi} \left ( 1 + \Delta \right ) \,, \qquad \Delta = \frac{\alpha_s}{4 \pi} \, f \! \left (x_{G^{\hspace{0.2mm} \prime}/U} \right ) \,, 
\end{equation}
for the partial decay width. Here $x_{G^{\hspace{0.2mm} \prime}/U} = M_{G^{\hspace{0.2mm} \prime}}^2/M_U^2$ and we have neglected the masses of the final state SM fermions. The function $f(x)$ that enters the partial decay width~(\ref{eq:GammaUbtau}) and encodes the ${\cal O} (\alpha_s)$ corrections takes the following form 
\begin{equation} \label{eq:fx}
\begin{split}
f(x) & = -\frac{4}{9} \left(7 x^2-27 x-37\right)-\frac{16 \hspace{0.125mm} \pi ^2}{9} +\frac{2}{9} \left(7 x^3-36 \hspace{0.125mm} x^2+21 x+30\right) \ln x \\[2mm]
& \phantom{xx} - \frac{4}{9} \left(7 x^2-22 \hspace{0.125mm} x-9\right) B(x) -\frac{16}{3} \left (2 x+1 \right ) C (x) \,,
\end{split}
\end{equation}
with 
\begin{equation} \label{eq:BxCx}
\begin{split} 
B(x) & = \sqrt{ \left(x-4\right) x} \hspace{0.25mm} \ln \left[ \frac{x+\sqrt{ \left(x-4\right) x}}{2 \sqrt{x}}\right] \,, \\[2mm]
C(x) & = -\frac{\pi ^2}{6} -\frac{1}{2} \hspace{0.125mm} \ln^2\left[\frac{\sqrt{ \left(x-4\right) x}-x}{2-x+\sqrt{ \left(x-4\right) x}}\right] \\[1mm]
& \phantom{xx} + \text{Li}_2\left[ \frac{2}{x + \sqrt{ \left(x-4\right) x}}\right] -\text{Li}_2\left [ \frac{2}{2 -x+\sqrt{ \left(x-4\right) x}}\right ]\,,
\end{split}
\end{equation}
where $\text{Li}_2 (z)$ is the usual dilogarithm. In the limit of degenerate singlet~vector~LQ and coloron masses it follows from~(\ref{eq:fx}) that $f (1) = 76/3- 32 \hspace{0.25mm} \pi/(3 \hspace{0.125mm} \sqrt{3})$, which coincides with the analytic expression reported in~\cite{Fuentes-Martin:2020luw}. This agreement serves as an independent cross check of the ${\cal O} (\alpha_s)$ calculations performed in the latter article. Notice that in the more generic case of the LQ-quark-lepton interactions~(\ref{eq:U1_fermion_interactions}) the total decay width of the LQ includes the processes $U \to b \tau$ and $U \to t \nu_\tau$, and can be obtained from~(\ref{eq:GammaUbtau}) by the simple replacement $g_4^2 \to g_U^2 \left [ \left ( 2 - 3 \hspace{0.125mm} x_{t/U}/2 +x_{t/U}^3/2 \right ) \left |\beta_L^{33} \right |^2 + \left |\beta_R^{33} \right |^2 \right ]/2$. Here $x_{t/U} = m_t^2/M_U^2$ and we have included the  corrections due to the non-negligible top-quark mass $m_t \simeq 163 \, {\rm GeV}$ that arise from the tree-level phase space and the squared matrix element at LO. Top-quark mass terms that arise at ${\cal O} (\alpha_s)$ and that would lead to a flavour-dependent correction $\Delta$ are instead neglected.  We believe this simplification to be an excellent approximation for LQ and coloron masses in the TeV range. Before moving on, let us finally add that the finite, renormalisation scale independent corrections~(\ref{eq:GammaUbtau}) also appear as universal ${\cal O} (\alpha_s)$ contributions to all low-energy observables that involve a LQ-quark-lepton vertex resulting from~(\ref{eq:4321_gauge_fermion}).  These corrections can be simply included by using,  instead of the tree-level coupling $g_4$,  the QCD corrected on-shell coupling $g_4 \left ( 1 + \Delta/2 \right)$ in the low-energy predictions~\cite{Fuentes-Martin:2020luw}.

\begin{figure}[t!]
\begin{center}
\includegraphics[width=0.475\textwidth]{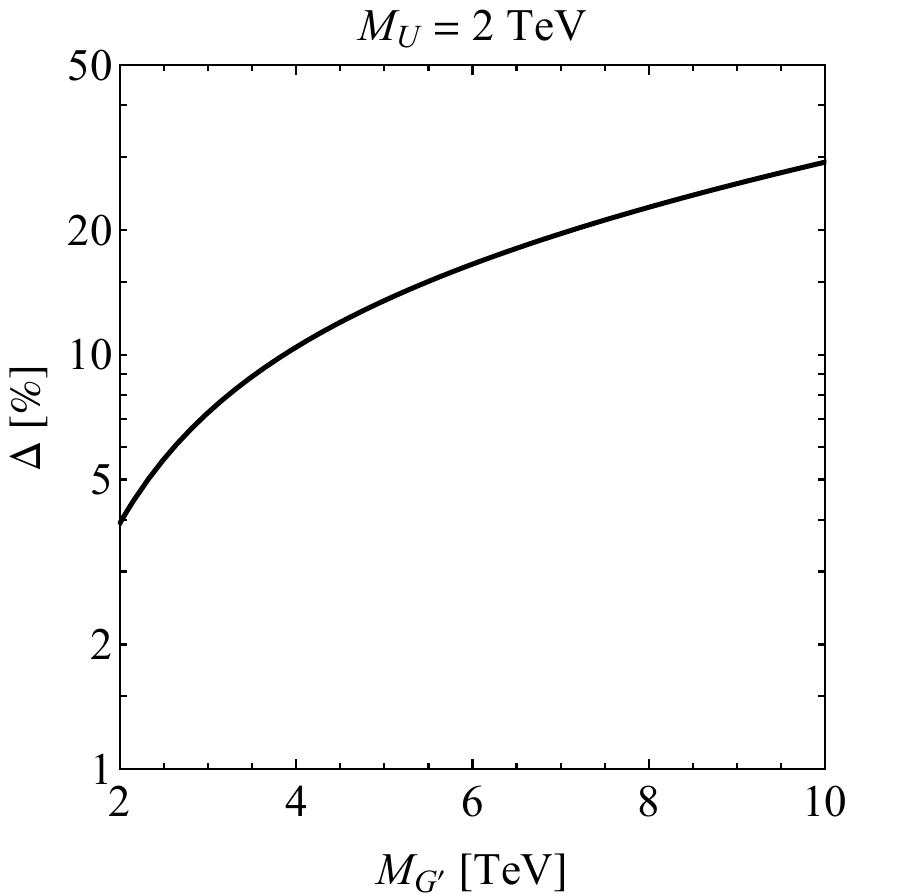}
\end{center}
\vspace{-4mm} 
\caption{\label{fig:width} Numerical size of the ${\cal O} (\alpha_s)$ correction to the partial decay width $U \to b \tau$ as~a~function~of the coloron mass~$M_{G^{\hspace{0.2mm} \prime}}$, fixing the singlet~vector~LQ mass to $M_U = 2 \, {\rm TeV}$. See~main text for~further~details.}
\end{figure}

In Figure~\ref{fig:width} we display the numerical size~of the NLO~QCD~correction $\Delta$ as defined in~(\ref{eq:GammaUbtau}). In the plot the mass of the singlet~vector~LQ is set to $M_U = 2 \, {\rm TeV}$. One~observes that the~${\cal O} (\alpha_s)$ corrections to the partial~decay~width~$U \to b \tau$ grow with increasing coloron mass~$M_{G^{\hspace{0.2mm} \prime}}$. For~$M_{G^{\hspace{0.2mm} \prime}} = 2 \, {\rm TeV}$,~$M_{G^{\hspace{0.2mm} \prime}} = 5 \, {\rm TeV}$ and~$M_{G^{\hspace{0.2mm} \prime}} = 10 \, {\rm TeV}$, we find that the NLO~QCD corrections amount to around 4\%, 14\% and 30\%, respectively. Notice that the observed enhancement originates from logarithmic non-decoupling contributions of the form~$\ln \left ( M_{G^{\hspace{0.2mm} \prime}}^2/M_U^2 \right)$. See~\cite{Fuentes-Martin:2019ign,Fuentes-Martin:2020luw} for detailed discussions of this issue. To gauge the ambiguities in our numerical analysis that are related to the choice of the masses of the heavy coloured vector states of the 4321 model, we will employ two benchmarks, namely~$M_{G^{\hspace{0.2mm} \prime}} = M_U$ and~$M_{G^{\hspace{0.2mm} \prime}} = 2.5 \hspace{0.2mm} M_U$. While the former choice is motivated by simplicity, the second option reflects the fact that the existing LHC bounds on the mass of the coloron are more stringent than those on the singlet~vector~LQ by at least a factor of two~\cite{Cornella:2021sby,ATLAS:2020dsk}. 

\begin{figure}[t!]
\begin{center} 
\includegraphics[width=0.475\textwidth]{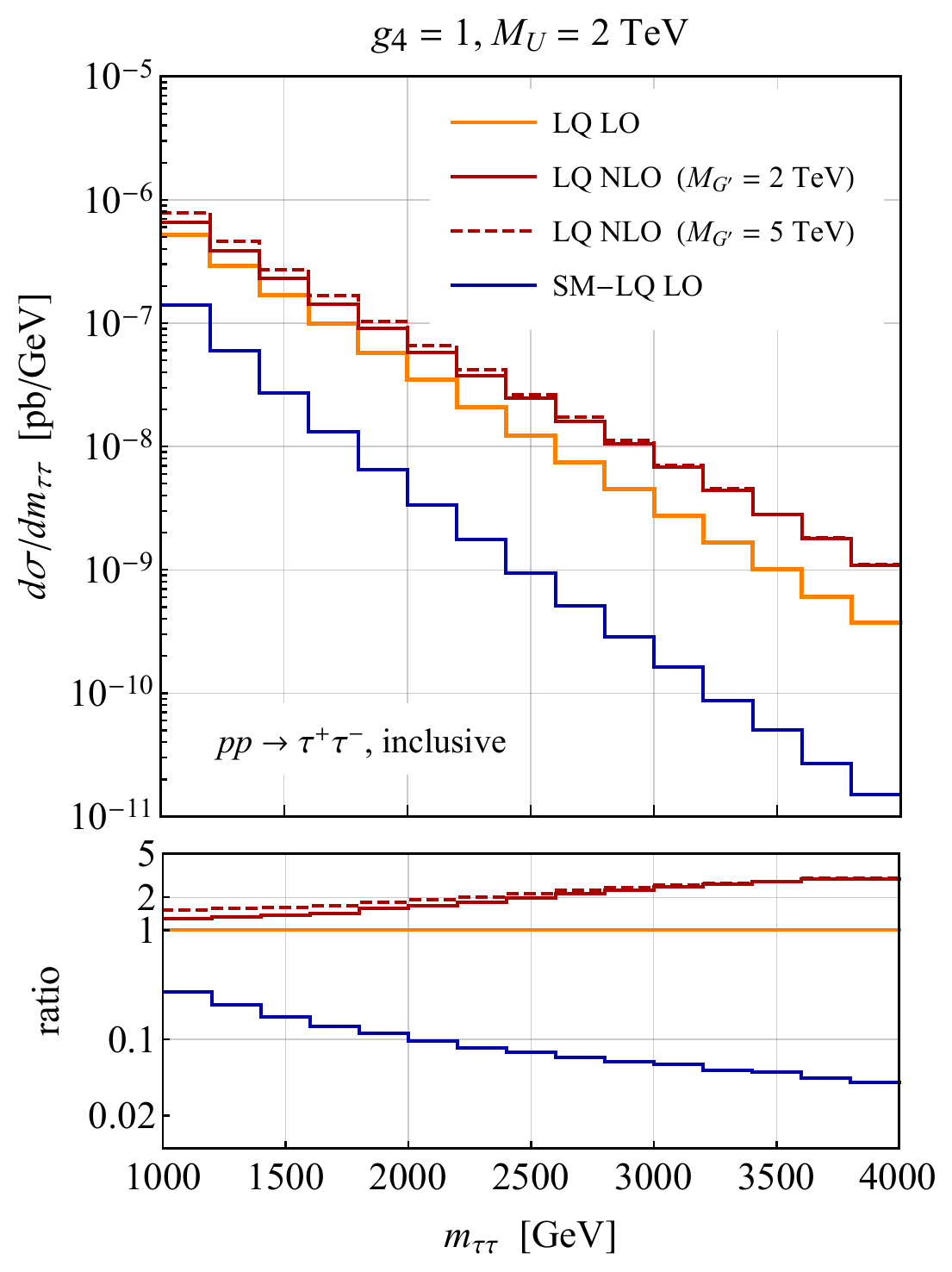} 
\end{center}
\vspace{-4mm} 
\caption{\label{fig:pheno1} Inclusive $pp \to \tau^+ \tau^-$ production cross sections as a function of~$m_{\tau \tau}$ for the parameter choices $g_4= 1$ and $M_U= 2 \, {\rm TeV}$. The yellow and red curves correspond to the~LQ distributions at the LO~(LQ~LO) and the~NLO~(LQ~NLO) in QCD, respectively, while the blue histograms illustrate the magnitude of the interference effects between the SM background and the LQ~signal~(SM-LQ~LO). In the case of the solid (dashed) red line the coloron mass is set to~$M_{G^{\hspace{0.2mm} \prime}} = 2 \, {\rm TeV}$ ($M_{G^{\hspace{0.2mm} \prime}} = 5 \, {\rm TeV}$). The~lower panel depicts the ratios between the different~LQ contributions and the relevant~LQ~LO~distribution.} 
\end{figure}

\begin{figure}[t!]
\begin{center} 
\includegraphics[width=0.475\textwidth]{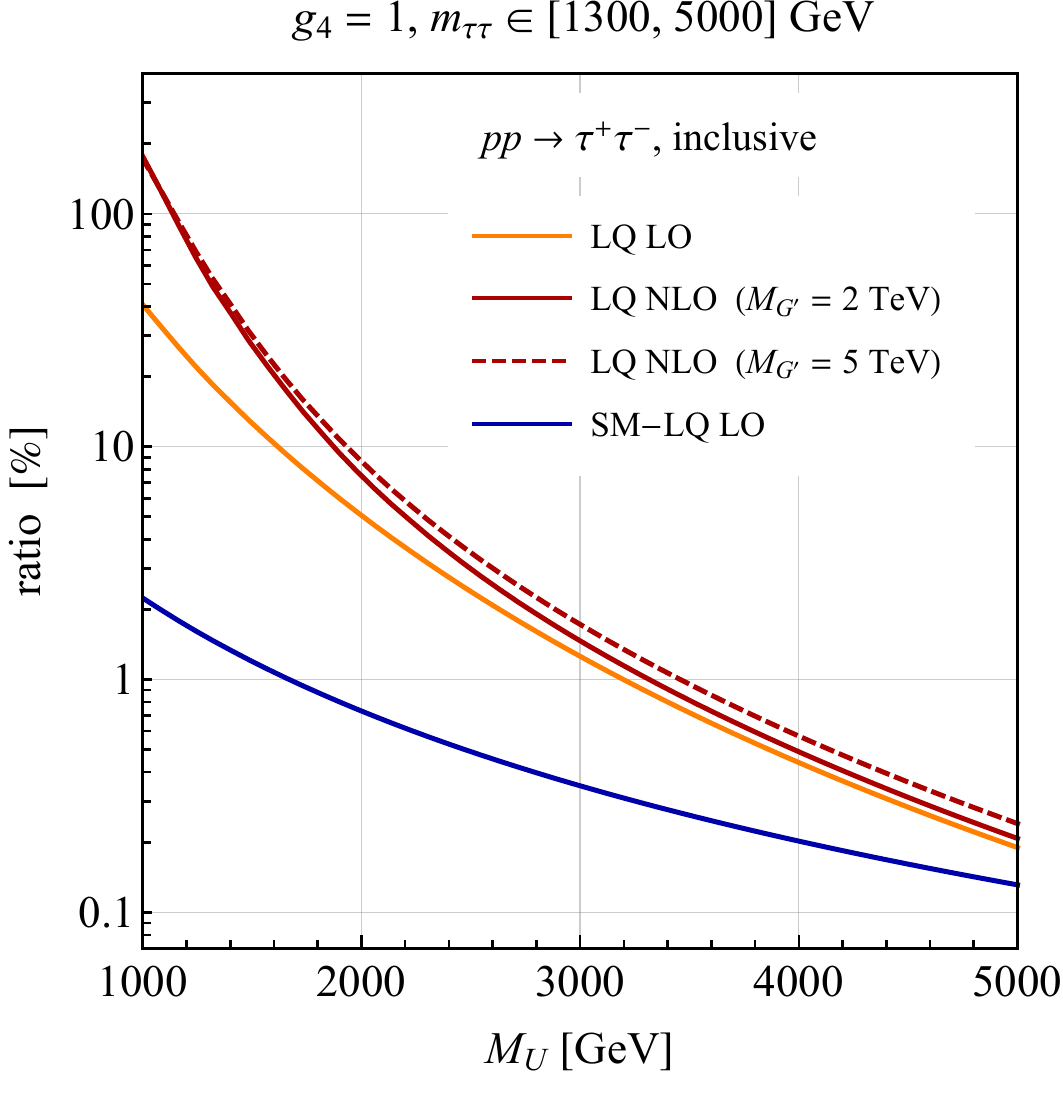} \quad 
\includegraphics[width=0.475\textwidth]{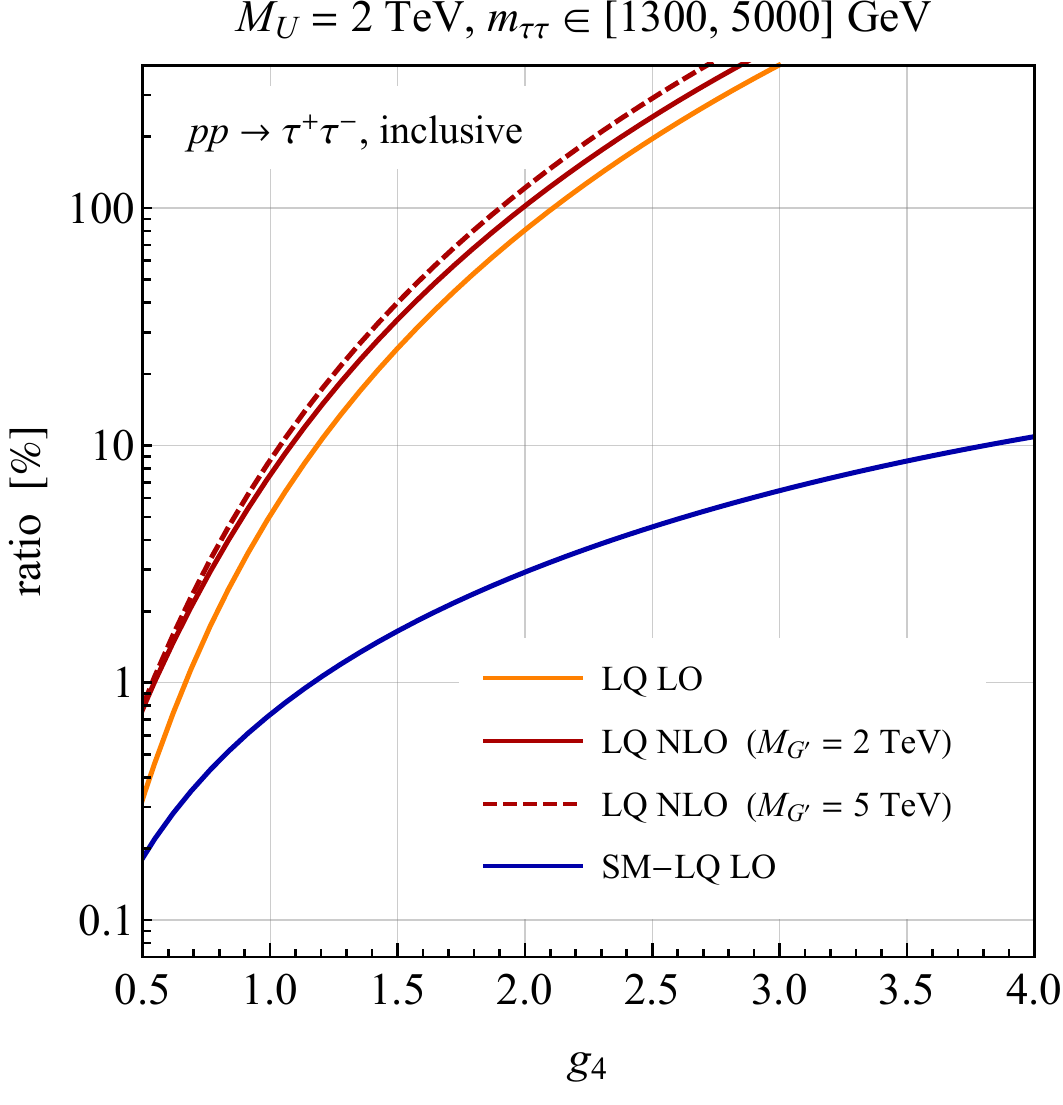}
\end{center}
\vspace{-4mm} 
\caption{\label{fig:pheno2} Ratios between the individual~LQ corrections and the~inclusive DY~SM background calculated at the NLO in QCD. The shown results correspond to the fiducial region defined by $p_{T, \hspace{0.25mm} \tau} > 30 \, {\rm GeV}$, $|\eta_\tau| < 2.5$ and~$m_{\tau \tau} \in [1300, 5000] \, {\rm GeV}$. The~left (right) plot depicts the results as a function of $M_U$ ($g_4$) for fixed $g_4 =1$ ($M_U = 2 \, {\rm TeV}$). The~colour coding and meaning of the different curves resembles those in Figure~\ref{fig:pheno1}. Additional details can be found in the main text.} 
\end{figure}

The simplest observable that one can study in DY ditau production is the invariant mass~$m_{\tau \tau}$ of the ditau system. In~Figure~\ref{fig:pheno1} we present our results for the~LQ corrections to the corresponding spectrum in inclusive $pp \to \tau^+ \tau^-$ production, employing {\tt NNPDF40\_nlo\_as\_01180} PDFs~\cite{Ball:2021leu}. The yellow and red lines resemble the~LQ distributions at the LO~(LQ~LO) and the~NLO~(LQ~NLO) in~QCD, respectively, while the blue curve illustrates the size of the interference effects between the SM background and the LQ~signature~(SM-LQ~LO). In the case of the solid~(dashed) red line the coloron mass is set to~$M_{G^{\hspace{0.2mm} \prime}} = 2 \, {\rm TeV}$ ($M_{G^{\hspace{0.2mm} \prime}} = 5 \, {\rm TeV}$). From~the lower panel of the plot it is evident that the NLO~QCD effects play an important role in obtaining precise predictions as they amount compared to the tree-level~LQ prediction to around~40\%~(150\%) at~$m_{\tau \tau} = 1.5 \, {\rm TeV}$ ($m_{\tau \tau} = 3 \, {\rm TeV}$). Notice that at NLO~in~QCD the DY ditau production spectra resulting from LQ exchange depend on the mass~$M_{G^{\hspace{0.2mm} \prime}}$ of the coloron. For the two choices of~$M_{G^{\hspace{0.2mm} \prime}}$ shown in the figure we find relative differences of the order of 10\% between the two distributions. The observed effects are therefore similar in size to the~$M_{G^{\hspace{0.2mm} \prime}}$ dependence of the ${\cal O} (\alpha_s)$ corrections to the partial decay width of the $U \to b \tau$ channel~(cf.~Figure~\ref{fig:width}). The interference effects between the SM DY background and the~LQ signal turn out to be destructive in the shown~$m_{\tau \tau}$ range,\footnote{The SM-LQ LO results shown in Figures~\ref{fig:pheno1}, \ref{fig:pheno2} and \ref{fig:pheno3} represent the magnitudes of the corresponding predictions for the interference effects between the SM background and the~LQ signal. } amounting to approximately 15\% (5\%) for~$m_{\tau \tau} = 1.5 \, {\rm TeV}$~($m_{\tau \tau} = 3 \, {\rm TeV}$). 

In~Figure~\ref{fig:pheno2} we furthermore display the ratios between the individual~LQ contributions and the DY~ditau~SM~background. The~normalisation is calculated at the NLO~in~QCD and we select events with two opposite-sign same-flavour tau leptons that are both required to have a transverse momentum of $p_{T, \hspace{0.25mm} \tau} > 30 \, {\rm GeV}$ and a pseudorapidity of $|\eta_\tau| < 2.5$. The~invariant masses of the ditau pairs must fall into the range~$m_{\tau \tau} \in [1300, 5000] \, {\rm GeV}$. Detector efficiency corrections are not taken into account. The~left panel displays our results as a function of $M_U$ fixing the overall coupling strength that appears in~(\ref{eq:4321_gauge_fermion}) to $g_4 = 1$. From this figure it is clearly visible that the relative size of the NLO~QCD corrections decreases for increasing~singlet vector LQ mass. Numerically, we find relative effects of around 330\%, 50\% and 15\% for $M_ U = 1 \, {\rm TeV}$, $M_U= 2 \, {\rm TeV}$ and $M_U = 3 \, {\rm TeV}$, respectively. This feature can be traced to the fact that the NLO~QCD corrections related to $s$-channel single-LQ production followed by the decay of the~LQ,~cf.~the~right Feynman diagram in Figure~\ref{fig:diagrams1}, decouple faster than the real and virtual corrections to the $t$-channel Born-level~LQ contribution,~cf.~the middle graph in Figure~\ref{fig:diagrams1} and the gluon-exchange diagrams in Figure~\ref{fig:diagrams2}. One~also observes that the interference effects represent only subleading corrections in the mass window~$m_{\tau \tau} \in [1300, 5000] \, {\rm GeV}$, amounting to an effect of at most~$-2\%$ relative to the SM background for the considered $M_U$ values. 

On the right-hand side in Figure~\ref{fig:pheno2} we finally depict our ratio predictions as a function of~$g_4$ setting the mass of the singlet vector LQ to $M_U = 2 \, {\rm TeV}$. It is evident from the plot that the relative size of the NLO~QCD corrections decreases for increasing overall coupling strength. In the case of~$M_{G^{\hspace{0.2mm} \prime}} = 2 \, {\rm TeV}$ the higher-order QCD effects amount compared to the tree-level~LQ result to around 140\%, 50\% and 30\% for $g_4=0.5$, $g_4=1$ and $g_4=2$. For~$M_{G^{\hspace{0.2mm} \prime}} = 5 \, {\rm TeV}$ the corresponding numbers read 150\%, 70\% and 50\%. This behaviour can be understood by realising that the~squared amplitude of the $t$-channel Born-level~LQ~contribution scales as $|g_4|^4$, while the resonant single-LQ production rate is proportional to~$|g_4|^2$. 
One~again sees that the interference contributions are numerically subleading even for large couplings~$g_4$ where they just reach the level of~$-10\%$. 

\section{Phenomenological analysis}
\label{sec:limits} 

LHC searches for signatures involving tau pairs in the final state like those performed in the publications~\cite{ATLAS:2020zms,CMS:2022goy,CMS-PAS-EXO-19-016} are known~\cite{Faroughy:2016osc,Schmaltz:2018nls,Baker:2019sli,Angelescu:2021lln,Bhaskar:2021pml,Cornella:2021sby} to provide strong constraints on~LQ~models that address the observed deviations in the charged-current $b \to c$ transitions. To illustrate the role that additional $b\hspace{0.4mm}$-jets play in analyses of this kind, we will consider as an example the recent~CMS~search~\cite{CMS:2022goy} for $\tau^+ \tau^-$ final states with both taus decaying to hadrons $($$\tau_{\rm h}$$)$. These $\tau_{\rm h}$ candidates are distinguished from jets originating from the hadronisation of light-flavoured quarks or gluons, and from electrons or muons by employing the $\tau$-tagger described in the~article~\cite{CMS:2022prd}. The used working points have an efficiency of approximately 50\%, 70\% and 70\% for identification in the case of jets, electrons and muons, respectively. The corresponding rejection factors are about 230, 20, and 770. Both~$\tau_{\rm h}$ candidates are required to have $p_{T, \tau} > 40 \, {\rm GeV}$ and $|\eta_{\tau}| < 2.1$, and their pseudorapidity-azimuth separation must be greater than $\Delta R_{\tau \tau} = 0.3$. Jets are clustered using the anti-$k_t$ algorithm with radius $R=0.4$, as implemented in~{\tt FastJet}~\cite{Cacciari:2011ma}. Light-flavoured quark or gluon jets need to fulfil $p_{T,j} > 30 \, {\rm GeV}$ and $|\eta_j| < 4.7$, while $b\hspace{0.4mm}$-jets with $p_{T,b} > 20 \, {\rm GeV}$ and $|\eta_b| < 2.5$ are selected. In order to identify $b\hspace{0.4mm}$-jets, we employ the~CMS~$b\hspace{0.4mm}$-tagging efficiencies stated in~\cite{CMS:2017wtu,Bols:2020bkb}. The used $b\hspace{0.4mm}$-tagging working point yields a~$b\hspace{0.4mm}$-tagging efficiency of around 80\% and a rejection in the ballpark of~100 for jets arising from light-flavoured quarks or gluons. Our~analysis is implemented into~{\tt MadAnalysis~5}~\cite{Conte:2012fm} and employs {\tt Delphes~3}~\cite{deFavereau:2013fsa} as a fast detector simulator. {\tt Pythia~8}~\cite{Sjostrand:2014zea} is used to shower the events. Effects~from hadronisation, underlying event modelling or QED effects in the PS are not included in our MC simulations. Applying~our~MC~chain to the SM~NLO~DY~prediction obtained with the {\tt POWHEG-BOX}, we are able reproduce the SM~DY~background as given~in~\cite{CMS:2022goy} to within around 30\%. This~comparison represents a non-trivial cross check of our ditau analysis.

In order to separate the LQ signal from the SM background, the distributions of the total transverse mass defined as~\cite{ATLAS:2014vhc}
\begin{equation} \label{eq:mTtot}
m_T^{\rm tot} = \sqrt{m_T^2 (\vec{p}_T^{\; \tau_1}, \vec{p}_T^{\; \tau_2}) + m_T^2 (\vec{p}_T^{\; \tau_1}, \vec{p}_T^{\; \rm miss}) + m_T^2 (\vec{p}_T^{\; \tau_2}, \vec{p}_T^{\; \rm miss}) } \,,
\end{equation}
are considered. Here $\tau_1$ ($\tau_2$) refers to the first (second) hadronic $\tau$ candidate and $\vec{p}_T^{\; \tau_1}$, $\vec{p}_T^{\; \tau_2}$ and $ \vec{p}_T^{\; \rm miss}$ are the vectors with magnitude $p_{T, \tau_1}$, $p_{T, \tau_2}$ and $E_{T ,\rm miss}$. The missing transverse energy constructed from the transverse momenta of all the neutrinos in the event is denoted by~$E_{T, \rm miss}$. The~transverse mass of two transverse momenta $p_{T,i}$ and $p_{T,j}$ entering~(\ref{eq:mTtot}) is given~by 
\begin{equation} \label{eq:mT}
m_T (\vec{p}_T^{\; i}, \vec{p}_T^{\; j} ) = \sqrt{2 \hspace{0.25mm} p_{T, i} \hspace{0.5mm} p_{T, j} \left ( 1 - \cos \Delta \phi \right )} \,, 
\end{equation}
where $\Delta \phi$ is the azimuthal angular difference between the vectors $\vec{p}_T^{\; i}$ and $ \vec{p}_T^{\; j}$.

\begin{figure}[t!]
\begin{center} 
\includegraphics[width=0.475\textwidth]{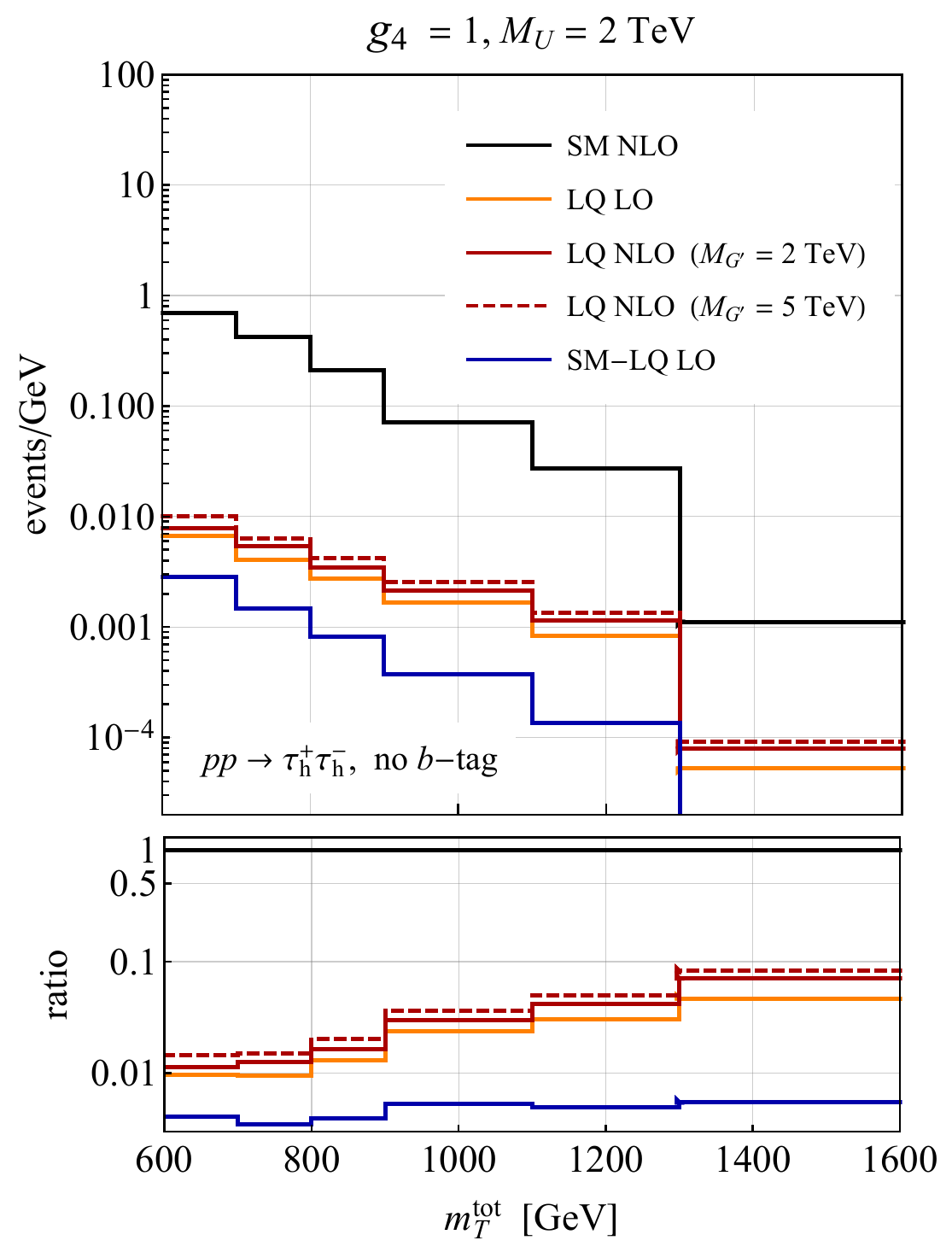} \quad 
\includegraphics[width=0.475\textwidth]{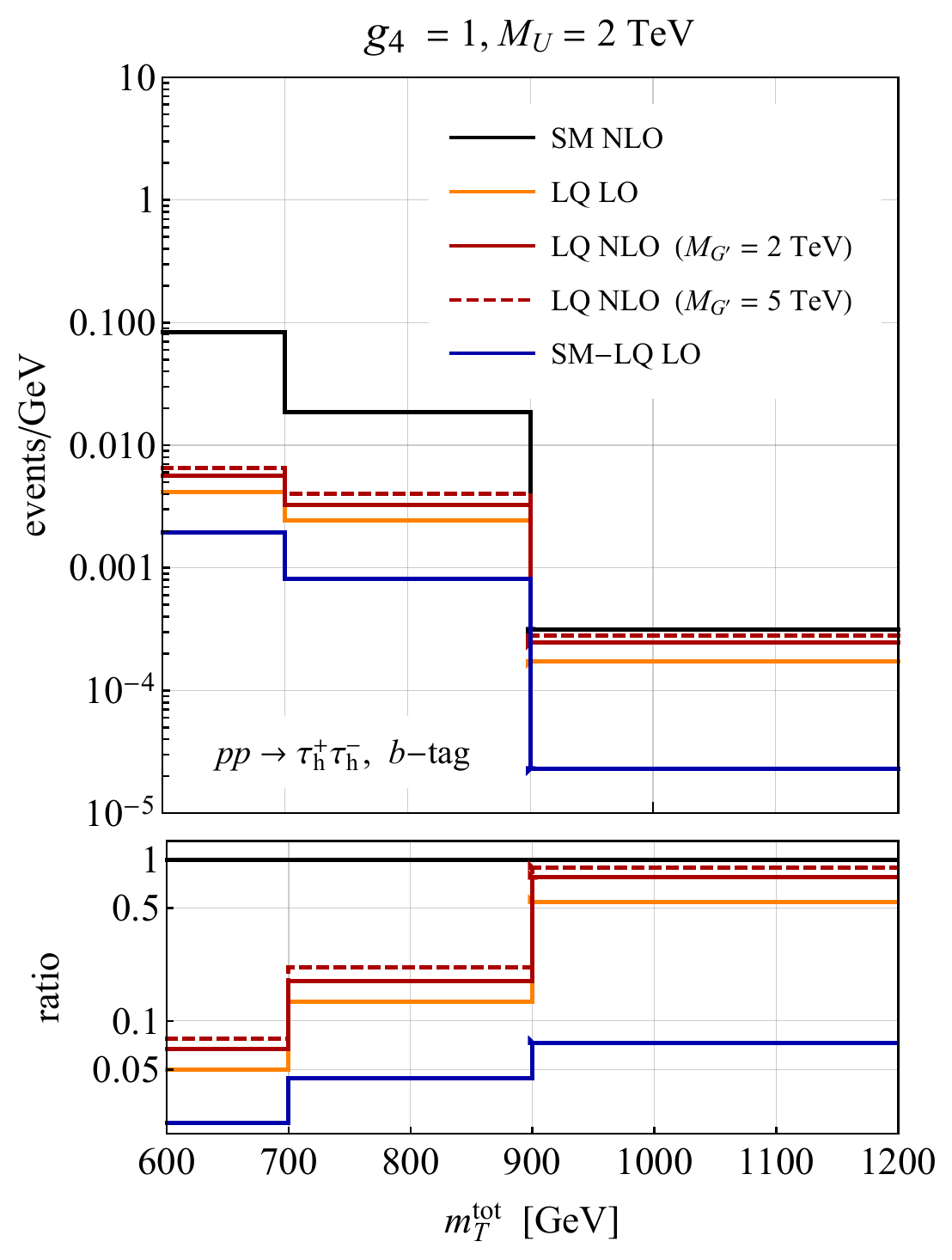}
\end{center}
\vspace{-4mm} 
\caption{\label{fig:pheno3} Distributions of $m_T^{\rm tot}$ in the no~$b\hspace{0.4mm}$-tag~(left~panel) and the~$b\hspace{0.4mm}$-tag (right~panel) category in the final state containing two hadronic tau leptons. The~black curves correspond to the SM expectations of the DY~background provided by~CMS in~\cite{CMS:2022goy}. This search is based on $138 \, {\rm fb}^{-1}$ of integrated luminosity collected in $pp$ collisions at $\sqrt{s} = 13 \, {\rm TeV}$. The yellow and red curves instead represent the~LQ~LO and~LQ~NLO predictions assuming $g_4 = 1$ and $M_U = 2 \, {\rm TeV}$. In~the case of the solid (dashed) red lines the coloron mass is set to~$M_{G^{\hspace{0.2mm} \prime}} = 2 \, {\rm TeV}$ ($M_{G^{\hspace{0.2mm} \prime}} = 5 \, {\rm TeV}$). The~blue histograms illustrate the size of the interference effects between the LQ signal and the SM~background called SM-LQ~LO. The~definition of the signal regions~(SRs) and other experimental details can be found in the main~text.}
\end{figure} 

In Figure~\ref{fig:pheno3} we compare the $m_T^{\rm tot}$ distributions as defined in~(\ref{eq:mTtot}) within the SM and the~4321~model~(\ref{eq:4321_gauge_fermion}) for the parameter choices $g_4 = 1$ and $M_U = 2 \, {\rm TeV}$. The~left~(right) panel displays the results for the no~$b\hspace{0.4mm}$-tag~($b\hspace{0.4mm}$-tag) category. The~black curves represent the SM expectations of the~DY~background taken from~\cite{CMS:2022goy}, while the yellow and red histograms are the~LQ~LO and~LQ~NLO predictions obtained using our {\tt POWHEG-BOX} code. The~solid~(dashed) red LQ~NLO results assume~$M_{G^{\hspace{0.2mm} \prime}} = 2 \, {\rm TeV}$ ($M_{G^{\hspace{0.2mm} \prime}} = 5 \, {\rm TeV}$). All predictions correspond to $138 \, {\rm fb}^{-1}$ of $pp$ data collected at $\sqrt{s} = 13 \, {\rm TeV}$. From the lower left panel one sees that in the no~$b\hspace{0.4mm}$-tag category the NLO~LQ contribution amounts to a relative correction of less than 10\% compared to the SM~DY~background for $m_T^{\rm tot} > 1300 \, {\rm GeV}$. For~what concerns the $b\hspace{0.4mm}$-tag category, one instead observes from the lower right panel that in the highest~$m_T^{\rm tot}$~bin with $m_T^{\rm tot} > 900 \, {\rm GeV}$ the NLO~LQ signal constitutes around 85\% of the SM~DY~background. This feature clearly shows that for third-generation vector~LQs the sensitivity of DY~searches notably improves by demanding an additional $b \hspace{0.4mm}$-jet in the final state. It is furthermore important to realise that the NLO~QCD effects enhance the~LO~LQ predictions in the no~$b\hspace{0.4mm}$-tag~($b\hspace{0.4mm}$-tag) category by approximately 35\% (30\%) in the highest~$m_T^{\rm tot}$~bin, making higher-order QCD effects phenomenologically relevant. On~the other hand, the dependence of the~NLO~LQ distributions on~$M_{G^{\hspace{0.2mm} \prime}}$ is weak. This renders the constraints derived below model-independent in the sense that one can set a limit on~$g_4$ as a function of~$M_U$ essentially without making a reference to the choice of the coloron mass~as long as~$M_{G^{\hspace{0.2mm} \prime}} = {\cal O} (M_U)$. One finally sees that the considered SM-LQ~LO~interference effects amount to a few permille in the case of the no~$b\hspace{0.4mm}$-tag category, while they can exceed the level of 5\% if one requires the presence of a $b\hspace{0.4mm}$-tag in the events. In contrast to what has been suggested in the recent work~\cite{CMS:2022goy}, interference effects therefore play only a minor role in the~SRs that are relevant for non-resonant~DY~searches for third-generation singlet vector~LQs at the LHC.

\begin{figure}[t!]
\begin{center} 
\includegraphics[width=0.475\textwidth]{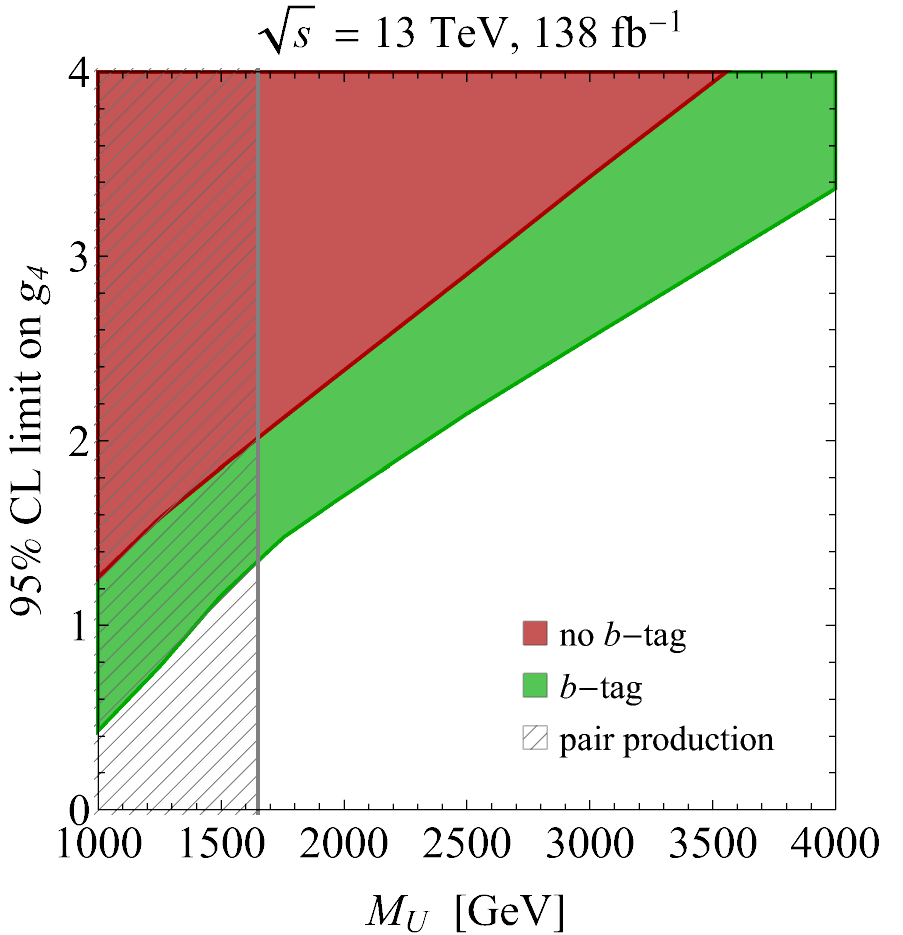} 
\end{center}
\vspace{-4mm} 
\caption{\label{fig:exclusion} Comparison of the 95\%~CL constraints on the $M_U\hspace{0.4mm}$--$\hspace{0.4mm} g_4$ plane that arise from the latest LHC~Run~II hadronic ditau analysis~\cite{CMS:2022goy}. The~red~(green) exclusion corresponds to the no~$b\hspace{0.4mm}$-tag~($b\hspace{0.4mm}$-tag) category of the latter search, while the hatched grey parameter space is excluded by strong pair production of third-generation~LQs~\cite{ATLAS:2021jyv}. Consult the main text for additional~explanations.} 
\end{figure}

Based on the ditau search strategies detailed above, we now derive NLO$+$PS accurate 95\%~confidence level~(CL) limits on the $M_U\hspace{0.4mm}$--$\hspace{0.4mm} g_4$ plane. Since we have seen that the choice of coloron mass has only a minor impact on the $m_T^{\rm tot}$ spectrum, we employ~$M_{G^{\hspace{0.2mm} \prime}} = M_U$ for simplicity when determining the exclusion bounds. Figure~\ref{fig:exclusion} shows our 95\%~CL limits on the $M_U\hspace{0.4mm}$--$\hspace{0.4mm} g_4$ parameter space that follow from the two $b\hspace{0.4mm}$-jet categories considered in the CMS search~\cite{CMS:2022goy} for two hadronic tau leptons. The~red and~green exclusion corresponds to the no~$b\hspace{0.4mm}$-tag and the~$b\hspace{0.4mm}$-tag category of this analysis, respectively, while the parameter space excluded by strong pair production~of~third-generation~LQs~\cite{CMS:2020wzx} is indicated by the hatched grey vertical band. This~search excludes $M_U < 1650 \, {\rm GeV}$ at 95\%~CL. The~significance of the individual $b\hspace{0.4mm}$-jet categories of the search~\cite{CMS:2022goy} is calculated as a ratio of Poisson likelihoods modified to incorporate systematic uncertainties on the background as Gaussian constraints~\cite{Cowan:2010js}. Our statistical analysis includes the six (three) highest $m_T^{\rm tot}$ bins in the case of the no~$b\hspace{0.4mm}$-tag ($b\hspace{0.4mm}$-tag) category. One first observes that the bound on $g_4$ that follows from the search with a $b\hspace{0.4mm}$-tag is more stringent than the one that derives from a strategy that requires no~$b\hspace{0.4mm}$-jet. 
We add that the difference between the no~$b\hspace{0.4mm}$-tag and $b\hspace{0.4mm}$-tag constraints is rather pronounced in the case of the CMS analysis~\cite{CMS:2022goy}, because this search observes a resonant-like excess with a significance of around $3\sigma$ at $m_T^{\rm tot} \simeq 1.2 \, {\rm TeV}$ in the no~$b\hspace{0.4mm}$-tag~sample. Consequently, the resulting no~$b\hspace{0.4mm}$-tag limits on the LQ parameter space are weaker than expected. Notice~finally that for $M_U \lesssim 1.7 \, {\rm TeV}$ the exclusions contour starts to~deviate from its linear behaviour. This is a consequence of the contribution associated to single-LQ production with subsequent decay of the~LQ,~cf.~the~right~diagram in~Figure~\ref{fig:diagrams1}, scaling as~$|g_4|^2$ compared to the~$|g_4|^4$ dependence of the~squared amplitude of the $t$-channel Born-level~LQ~contribution. 

\section{Conclusions}
\label{sec:conclusions}

The main goal of this article was to refine the theoretical description of DY~dilepton production in vector LQ models. To this purpose we have calculated the NLO~QCD corrections to the~$pp \to \ell^+ \ell^-$ process. The actual~computation involves the evaluation of the real and virtual corrections to the~$t$-channel Born-level contribution and the calculation of resonant single-LQ production followed by the decay of the~LQ. One complication compared to the computation of ${\cal O} (\alpha_s)$ corrections to DY dilepton production in scalar LQ models~\cite{Alves:2018krf,Haisch:2022lkt} arises from the fact that realistic vector~LQ models such as~the~4321~model~(\ref{eq:4321_gauge_fermion}) contain additional states that carry non-zero~$SU(3)_C$ charges. In~fact, in the case at hand both gluon and coloron exchange has to be considered in order to determine the full~NLO~QCD contributions to DY dilepton production. Our~${\cal O} (\alpha_s)$~computation furthermore serves as an independent cross check of the calculation of the singlet vector~LQ~decay width in the 4321 model presented in~\cite{Fuentes-Martin:2020luw}. Besides QCD corrections we have also studied the size of interference effects between the DY~SM background and the LQ signature, finding that these effects are in general small in the SRs of the existing~LHC~DY~dilepton~searches. 

The calculated fixed-order predictions have been implemented into a dedicated MC~code which consistently matches them to a PS employing the {\tt POWHEG} method. As a result, a~realistic exclusive description of DY~dilepton processes in the singlet vector~LQ~model at the level of hadronic events can be obtained without the introduction of an unphysical merging or matching scale. Our MC generator should prove useful for everyone interested in comparing accurate theory predictions to LHC data, and we therefore make the relevant code to simulate NLO+PS events for the $pp \to \ell^+ \ell^-$ process in singlet vector~LQ~model of the form~(\ref{eq:4321_gauge_fermion}) available for download on the official {\tt POWHEG-BOX}~web~page~\cite{POWHEGBOX}. 

In our phenomenological analysis, we have studied the case of $pp \to \tau^+ \tau^-$ production that arises from the LQ-quark-lepton couplings~(\ref{eq:4321_gauge_fermion}) supplemented by the pure gauge, Goldstone boson and ghost contributions entering~(\ref{eq:4321_other}). The~focus on ditau final states is motivated by the observation~\cite{Faroughy:2016osc,Schmaltz:2018nls,Baker:2019sli,Angelescu:2021lln,Bhaskar:2021pml,Cornella:2021sby,CMS:2022goy} that models providing an explanation to the charged-current~$b \to c$ anomalies in general also predict enhanced $pp \to \tau^+ \tau^-$ rates. Since these ditau signatures result from bottom-quark fusion, initial-state radiation will always lead to an enhanced~$b\hspace{0.4mm}$-jet activity in the events. Devising~search strategies with different $b\hspace{0.4mm}$-jet categories is therefore expected to help improve the LHC sensitivity~\cite{ATLAS:2020zms,ATLAS:2021mla,CMS:2022goy,Afik:2018nlr,Choudhury:2019ucz,Altmannshofer:2017poe,Iguro:2017ysu,Abdullah:2018ets,Marzocca:2020ueu,Endo:2021lhi,Haisch:2022lkt,CMS-PAS-EXO-19-016}. To illustrate this point, we have performed a recast of the search~\cite{CMS:2022goy} that employs $138 \, {\rm fb}^{-1}$ of~$pp$~data collected at $\sqrt{s} = 13 \, {\rm TeV}$. This analysis studies two disjoint SRs, and we found that the search strategy that requires the presence of an additional $b\hspace{0.4mm}$-tagged jet outperforms~the~search~strategy that vetos $b\hspace{0.25mm}$-jets. Utilising~\cite{CMS:2022goy} together with our {\tt POWHEG-BOX} implementation we have finally derived NLO$+$PS accurate constraints on the masses and couplings of the 4321 model~(\ref{eq:4321_gauge_fermion}). In~Appendix~\ref{app:moreconstraints}~we furthermore provide the constraints on the parameter space of third-generation singlet vector~LQs that arise from the LHC~Run~II analyses~\cite{ATLAS:2020zms,CMS-PAS-EXO-19-016} of ditau production. We~emphasise that the presented {\tt POWHEG-BOX} generator provides an improved signal modelling compared to the matched~MLM~\cite{Alwall:2007fs}~LO~{\tt MadGraph5\_aMC\@NLO}~\cite{Alwall:2014hca} samples used in~\cite{CMS:2022goy}. Similar statements also apply to the signal generations used in the analyses~\cite{ATLAS:2020zms,CMS-PAS-EXO-19-016}. This makes our~MC~implementation an essential tool for ATLAS and CMS searches for singlet vector~LQs~in ditau final states at future LHC runs. 

\acknowledgments{We thank Javier Fuentes-Mart{\'i}n for useful discussions and Benjamin~Fuks for his help regarding the expert mode of \texttt{MadAnalysis~5}.  The Feynman diagrams shown in this work have been drawn with~{\tt JaxoDraw}~\cite{Binosi:2008ig}. LS~and~SS are supported by the International Max Planck Research School (IMPRS) on “Elementary Particle Physics”. Partial support by the Collaborative Research Center SFB1258 is also acknowledged. UH~and~LS would like to express gratitude to the Mainz Institute for Theoretical Physics (MITP) of the Cluster of Excellence PRISMA+ (Project ID 39083149), for its hospitality and support in the initial stage of this project. We finally thank an unknown referee for raising the point about the possible impact of $Z^\prime$ exchange in $pp \to \tau^+ \tau^-$ production in the context of the 4321 model, which led to the study presented in Appendix~\ref{app:Zprime}.}
\begin{appendix}

\newpage

\section{Feynman rules}
\label{app:feynman}

To obtain the complete ${\cal O} (\alpha_s)$ contribution to DY~dilepton~production in the 4321 model, one has to consider besides the interactions (\ref{eq:4321_gauge_fermion}) also pure gauge, Goldstone boson and ghost contributions. The non-fermionic interaction Lagrangian necessary to perform the NLO~QCD calculation described in Section~\ref{sec:calculation} takes the form 
\begin{equation} \label{eq:4321_other}
\begin{split}
{\cal L}_{4321} & \supset \, i \hspace{0.2mm} g_s \left[ \left( U^\dagger_{\mu \nu} \hspace{0.2mm} G^{\mu, a} \hspace{0.2mm} T^a \hspace{0.2mm} U^\nu + {\rm h.c.} \right) -\hspace{0.2mm} U_\mu^\dagger \hspace{0.2mm} T^a \hspace{0.2mm} U_\nu \hspace{0.4mm} G^{\mu \nu, a} \right] \\[2mm]
& \phantom{xx} +i \hspace{0.2mm} c_3 \hspace{0.4mm} g_4 \left[\left(U^\dagger_{\mu \nu} G^{\hspace{0.2mm} \prime \hspace{0.2mm} \mu, a} T^a U^\nu + {\rm h.c.} \right) - \hspace{0.2mm} U_\mu^\dagger \hspace{0.2mm} T^a \hspace{0.2mm} U_\nu \hspace{0.4mm} G^{\hspace{0.2mm} \prime \hspace{0.2mm} \mu \nu, a} \right] \\[2mm]
& \phantom{xx} + g_s \hspace{0.4mm} M_U \hspace{0.4mm} U_\mu^\dagger \hspace{0.2mm} T^a \hspace{0.2mm} \pi_U \hspace{0.4mm} G^{\mu, a} + 
 c_3 \hspace{0.4mm} g_4 \hspace{0.4mm} \frac{M_U^2-M_{G^{\hspace{0.2mm} \prime}}^2}{M_U} \hspace{0.4mm} U_\mu^\dagger \hspace{0.2mm} T^a \hspace{0.2mm} \pi_U \hspace{0.4mm} G^{\hspace{0.2mm} \prime \hspace{0.2mm} \mu, a} + {\rm h.c.} \\[2mm]
 & \phantom{xx} +i \hspace{0.2mm} g_s \hspace{0.4mm} \Big[\left( \partial_\mu \bar{c}_{U} \right) \hspace{0.2mm} T^a \hspace{0.2mm} U^\mu \hspace{0.2mm} c_{G^a} -\hspace{0.2mm} U^\dagger_{\mu} \hspace{0.2mm} T^a \hspace{0.2mm} \left( \partial^\mu \bar{c}_{U^\dagger} \right) \hspace{0.2mm} c_{G^a} \\[2mm]
 & \phantom{xxxxxxx} - \hspace{0.2mm} \left( \partial_\mu \bar{c}_{G^a} \right) c_{U^\dagger} \hspace{0.2mm} T^a \hspace{0.2mm} U^\mu +\hspace{0.2mm} \left( \partial_\mu \bar{c}_{G^a} \right) U^{\dagger \hspace{0.2mm} \mu} \hspace{0.2mm} T^a \hspace{0.2mm} c_U \Big] \\[2mm]
 & \phantom{xx} +i \hspace{0.2mm} c_3 \hspace{0.4mm} g_4 \hspace{0.4mm} \Big[\left( \partial_\mu \bar{c}_{U} \right) \hspace{0.2mm} T^a \hspace{0.2mm} U^\mu \hspace{0.2mm} c_{G^{\prime \hspace{0.2mm} a}} -\hspace{0.2mm} U^\dagger_{\mu} \hspace{0.2mm} T^a \hspace{0.2mm} \left( \partial^\mu \bar{c}_{U^\dagger} \right) \hspace{0.2mm} c_{G^{\prime \hspace{0.2mm} a}} \\[2mm]
 & \phantom{xxxxxxx} - \hspace{0.2mm} \left( \partial_\mu \bar{c}_{G^{\prime \hspace{0.2mm} a}} \right) c_{U^\dagger} \hspace{0.2mm} T^a \hspace{0.2mm} U^\mu +\hspace{0.2mm} \left( \partial_\mu \bar{c}_{G^{\prime \hspace{0.2mm} a}} \right) U^{\dagger \hspace{0.2mm} \mu} \hspace{0.2mm} T^a \hspace{0.2mm} c_U \Big] \,.
\end{split}
\end{equation}
Here $X_{\mu \nu} = \partial_\mu X_\nu - \partial_\nu X_\mu$ for $X = U, U^\dagger, G^{a}, G^{\prime \hspace{0.2mm} a}$ are the relevant field strength tensors, $\pi_U$~is the Goldstone boson associated with the radial polarisation of the singlet vector~LQ and~$c_X$~are the ghost fields originating from the Fadeev-Popov gauge fixing procedure applied to the gauge boson field $X$.

\section{Further constraints}
\label{app:moreconstraints}

In this appendix we present the 95\%~CL limits on the $M_U\hspace{0.4mm}$--$\hspace{0.4mm} g_4$ plane that follow from recasts of the LHC~Run~II analyses~\cite{ATLAS:2020zms,CMS-PAS-EXO-19-016} of ditau production. The event generation is again performed at the NLO$+$PS level using the~{\tt POWHEG-BOX} implementation described in the main part of this work. We use {\tt NNPDF40\_nlo\_as\_01180} PDFs, {\tt Pythia~8} as a PS and {\tt MadAnalysis~5} together with {\tt Delphes~3} as an analysis tool. As before underlying event modelling or QED effects in the PS are not included in our MC simulations. Applying~our~MC~chain to the SM prediction for $pp \to \tau^+ \tau^-$ obtained with the {\tt POWHEG-BOX} at NLO$+$PS, we are able reproduce the relevant SM~DY~background distributions as given~in~\cite{ATLAS:2020zms,CMS-PAS-EXO-19-016} to about 30\%. This approximate agreement serves as an important cross check of our analysis framework. 

The search strategy for hadronic tau leptons used by ATLAS~in~\cite{ATLAS:2020zms} is quite similar to that of CMS as described in~\cite{CMS:2022goy}. The~hadronic~$\tau$ candidates are composed of a neutrino and a set of visible decay products~($\tau_{\text{had-vis}}$), usually consisting of one or three charged pions and up to two neutral pions. These~$\tau_{\text{had-vis}}$ candidates are reconstructed from seeding jets~\cite{ATLAS-CONF-2017-029} and are required to have $p_{T, \tau} > 65 \, {\rm GeV}$ and $|\eta_{\tau}|<2.5$. The~$\tau_{\text{had-vis}}$ candidates must satisfy loose or medium~$\tau$ identification criteria with efficiencies of about 85\%~(75\%) and 75\%~(60\%) for one-track~(three-track) candidates, respectively. The corresponding rejections factors in multijet events are roughly 20~(200) and 30~(500) for one-track (three-track) candidates~\cite{ATLAS-CONF-2017-029}. The~two~hadronic~$\tau$ candidates are required to have opposite electric charge and the azimuthal angular difference between the vectors $\vec{p}_T^{\; \tau_1}$ and $ \vec{p}_T^{\; \tau_2}$ needs to fulfil $|\Delta \phi| > 2.7$. Jets are clustered using the anti-$k_t$ algorithm with radius $R=0.4$ and must satisfy $p_{T,j} > 20 \, {\rm GeV}$ and $|\eta_j| < 2.5$. Our $b\hspace{0.4mm}$-jet identification is based on the information provided in the ATLAS note~\cite{ATLAS:2019bwq}. The~used~$b\hspace{0.4mm}$-tagging working point yields a~$b\hspace{0.4mm}$-tagging efficiency of around 70\% and rejections of approximately~9,~36~and~300 for $c$-jets, $\tau$ decays involving hadrons and jets arising from light-flavoured quarks or gluons, respectively. Like in the case of the CMS analysis~\cite{CMS:2022goy} the total transverse mass~(\ref{eq:mTtot}) is used in~\cite{ATLAS:2020zms} and our recast to discriminate between the LQ signal and the SM background. Two distinct SRs, one where $b\hspace{0.4mm}$-jets are vetoed and another one that require a $b\hspace{0.4mm}$-jet in the event, are then studied. 

\begin{figure}[t!]
\begin{center} 
\includegraphics[width=0.475\textwidth]{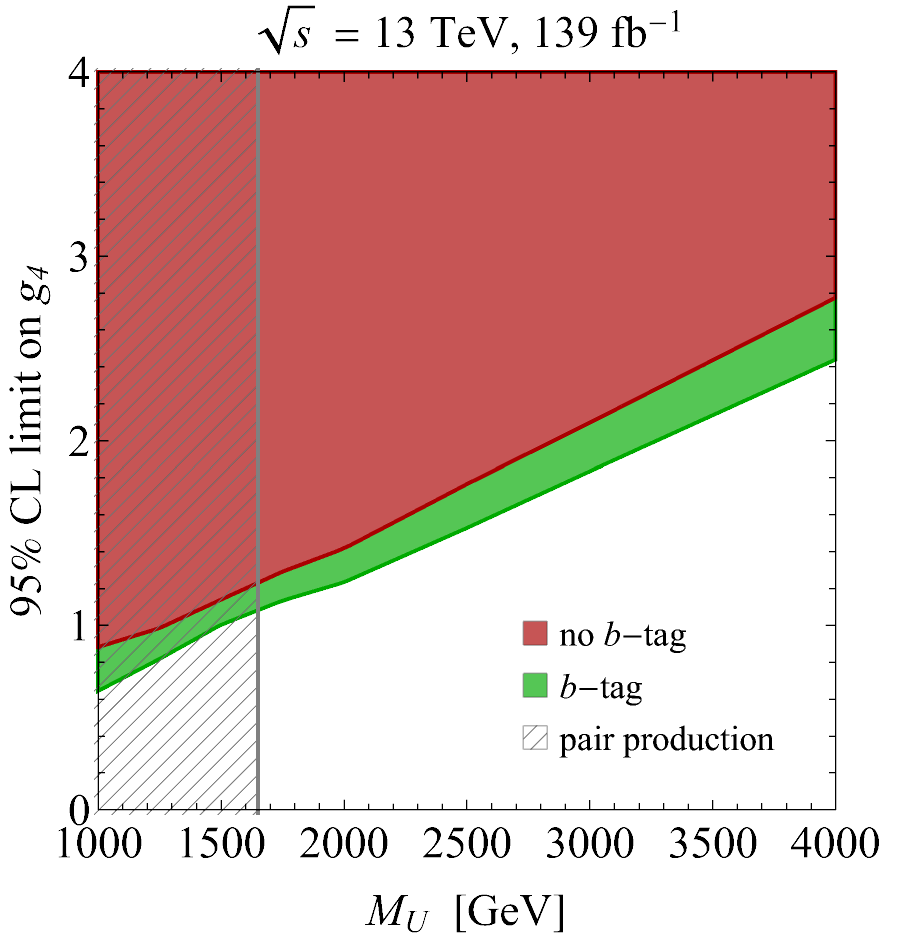} \quad 
\includegraphics[width=0.475\textwidth]{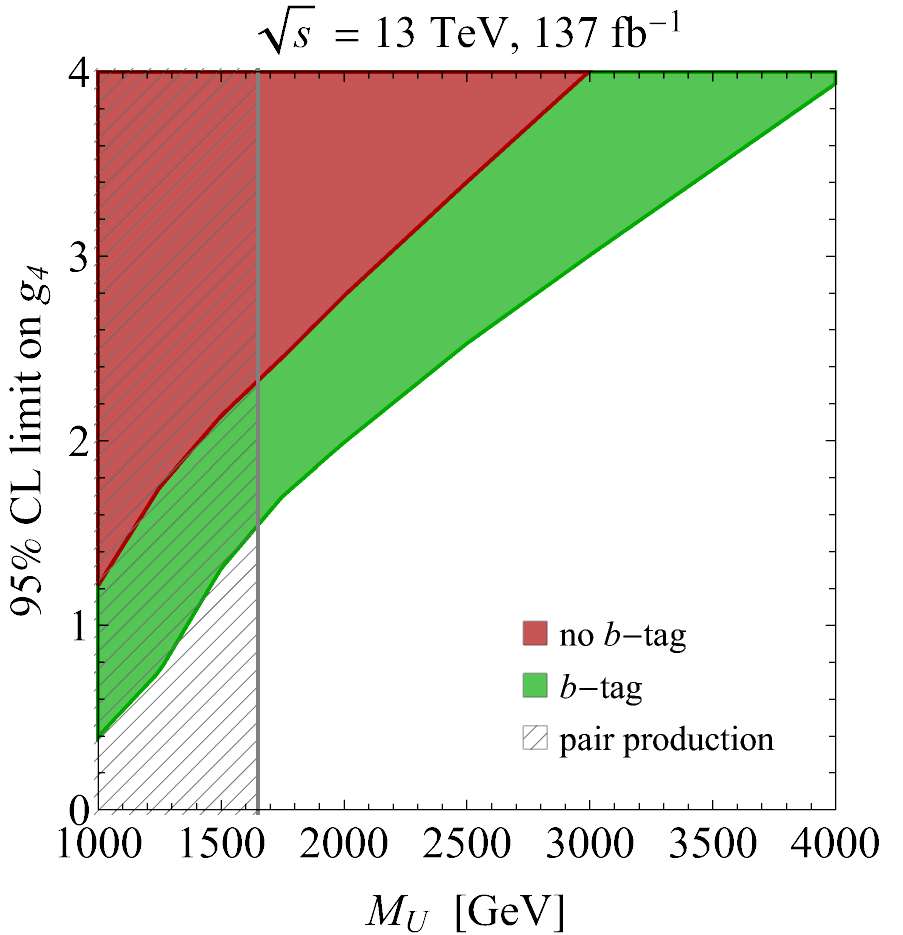} 
\end{center}
\vspace{-4mm} 
\caption{\label{fig:exclusion2} As Figure~\ref{fig:exclusion} but using a recast of the results of the ATLAS~\cite{ATLAS:2020zms} and CMS~\cite{CMS-PAS-EXO-19-016} ditau search in the~left and right panel, respectively. For additional details see the text. }
\end{figure}

The latest ditau search by CMS~\cite{CMS-PAS-EXO-19-016} instead imposes the following selection requirements. Events with two hadronic $\tau$ candidates with opposite-sign electric charge are selected. The $\tau_h$ candidates are reconstructed with the so-called hadron-plus-strips algorithm~\cite{CMS:2022prd,CMS:2018jrd}. The medium working point of this algorithm is used in our recast which has an efficiency of about 70\% for a genuine $\tau_h$ and a misidentification rate of around~0.1\% for light-flavoured quark or gluon jets. We furthermore require that $p_{T,\tau} > 50 \, {\rm GeV}$, $|\eta_\tau| < 2.3$ and $\Delta R_{\tau \tau} > 0.5$. Jets are clustered with the anti-$k_t$ algorithm and $R=0.4$. Our analysis selects all jets that satisfy $p_{T,j} > 50 \, {\rm GeV}$ and $|\eta_j| < 4.7$. The identification of $b\hspace{0.4mm}$-jets employs a parameterisation of the loose working point of~\cite{CMS:2017wtu,Guest:2016iqz}. The efficiency of this $b\hspace{0.4mm}$-tagger can reach up to 90\% but degrades down to approximately 60\% for $p_{T,b} > 500 \, {\rm GeV}$. To~remove DY background an~additional cut on the invariant mass~$m_{\rm vis}$ of the visible tau decay products of~$m_{\rm vis} > 100 \, {\rm GeV}$ is applied. The~scalar~sum 
\begin{equation} \label{eq;STMET}
S_T^{\rm MET} = p_{T, \tau_1} + p_{T, \tau_2} + p_{T,j} + E_{T, \rm miss} \,,
\end{equation}
built from the transverse momenta $p_{T, \tau_1}$ and $p_{T, \tau_2}$ of the two $\tau$ candidates, the transverse momentum $p_{T,j}$ of the leading jet and the missing transverse energy $E_{T, \rm miss}$ is used in the analysis~\cite{CMS-PAS-EXO-19-016} as a discriminating variable. Furthermore, two orthogonal event categories are constructed: one which requires no~$b\hspace{0.4mm}$-jet with $p_{T,b} > 50 \, {\rm GeV}$ and another one which requires at least one such jet. 

The 95\%~CL exclusion bounds on the $M_U\hspace{0.4mm}$--$\hspace{0.4mm} g_4$ plane that follow from the recast of the~ATLAS~\cite{ATLAS:2020zms} and~CMS~\cite{CMS-PAS-EXO-19-016} search are shown in the left and right panel of Figure~\ref{fig:exclusion2}, respectively. For simplicity we again employ~$M_{G^{\hspace{0.2mm} \prime}} = M_U$ when determining the exclusion limits. Compared to the constraints depicted in Figure~\ref{fig:exclusion}, one observes that the difference between the no~$b\hspace{0.4mm}$-tag and $b\hspace{0.4mm}$-tag bounds that derive from the considered ATLAS analysis is much smaller. This feature is readily understood by noticing that the ATLAS search, unlike the CMS analysis~\cite{CMS:2022goy} does not see an excess in the high-mass $m_T^{\rm tot}$ distribution in the no $b\hspace{0.4mm}$-tag category. In~fact, ATLAS observes small deficits compared to the expected SM background in the~tails of the~$m_T^{\rm tot}$ spectra, which explains why for large values of $M_U$ the 95\%~CL limits on~$g_4$ as shown in the left panel of~Figure~\ref{fig:exclusion2} are notably better than those displayed in~Figure~\ref{fig:exclusion}. To~understand the shape of the~exclusion limits following from the CMS~search~\cite{CMS-PAS-EXO-19-016} presented on the right-hand side in Figure~\ref{fig:exclusion2}, one has to realise that the latter search observes a non-resonant excess with a significance of a bit more than~$3\sigma$ above the SM expectation in the data. As~a~result, the obtained 95\%~CL limits in the $M_U\hspace{0.4mm}$--$\hspace{0.4mm} g_4$ plane turn out to be weaker than expected, in particular in the large mass regime. 

\section{Ditau production from $\bm{Z^\prime}$ exchange}
\label{app:Zprime}

\begin{figure}[t!]
\begin{center} 
\includegraphics[width=0.475\textwidth]{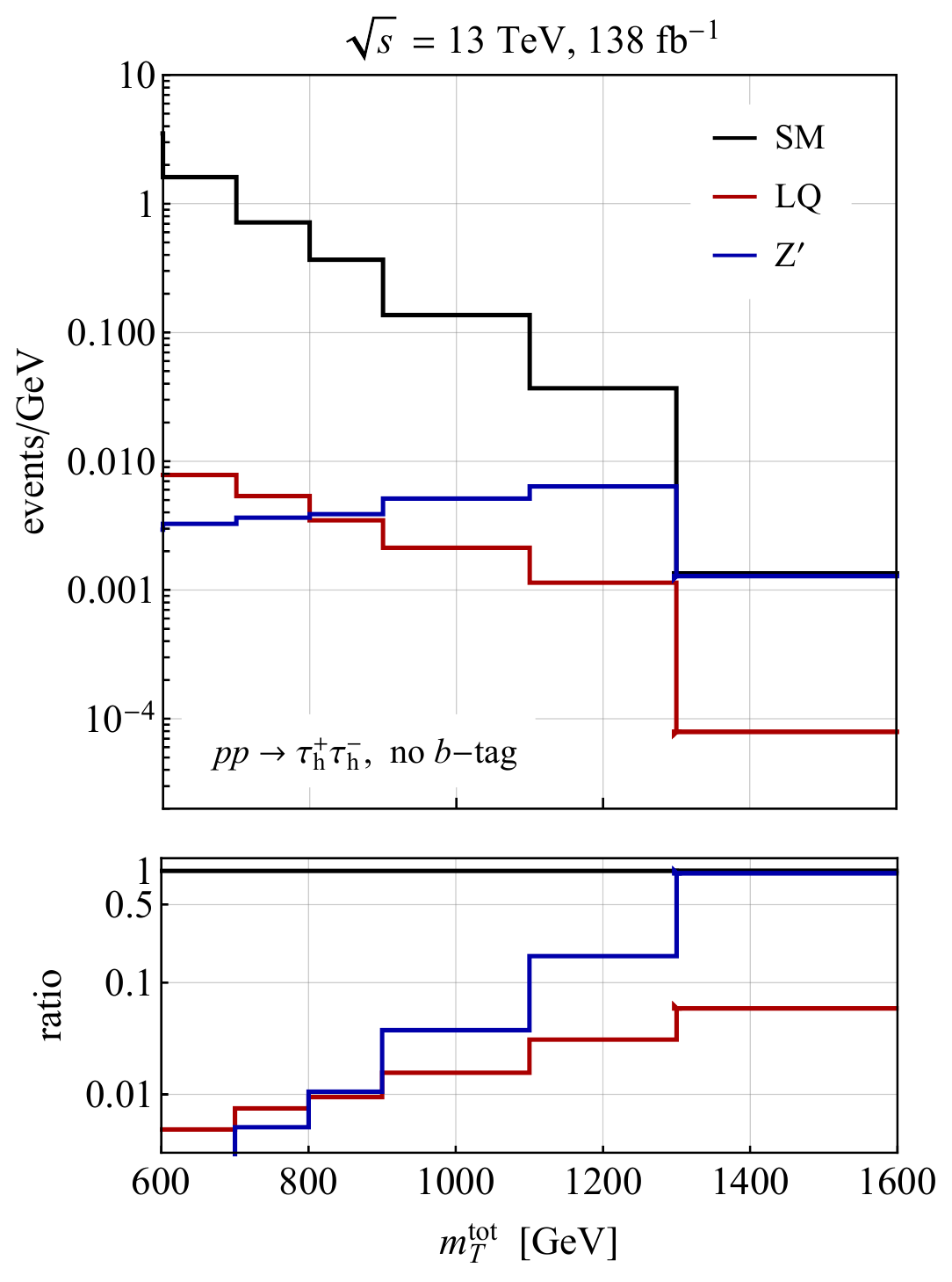} \quad 
\includegraphics[width=0.475\textwidth]{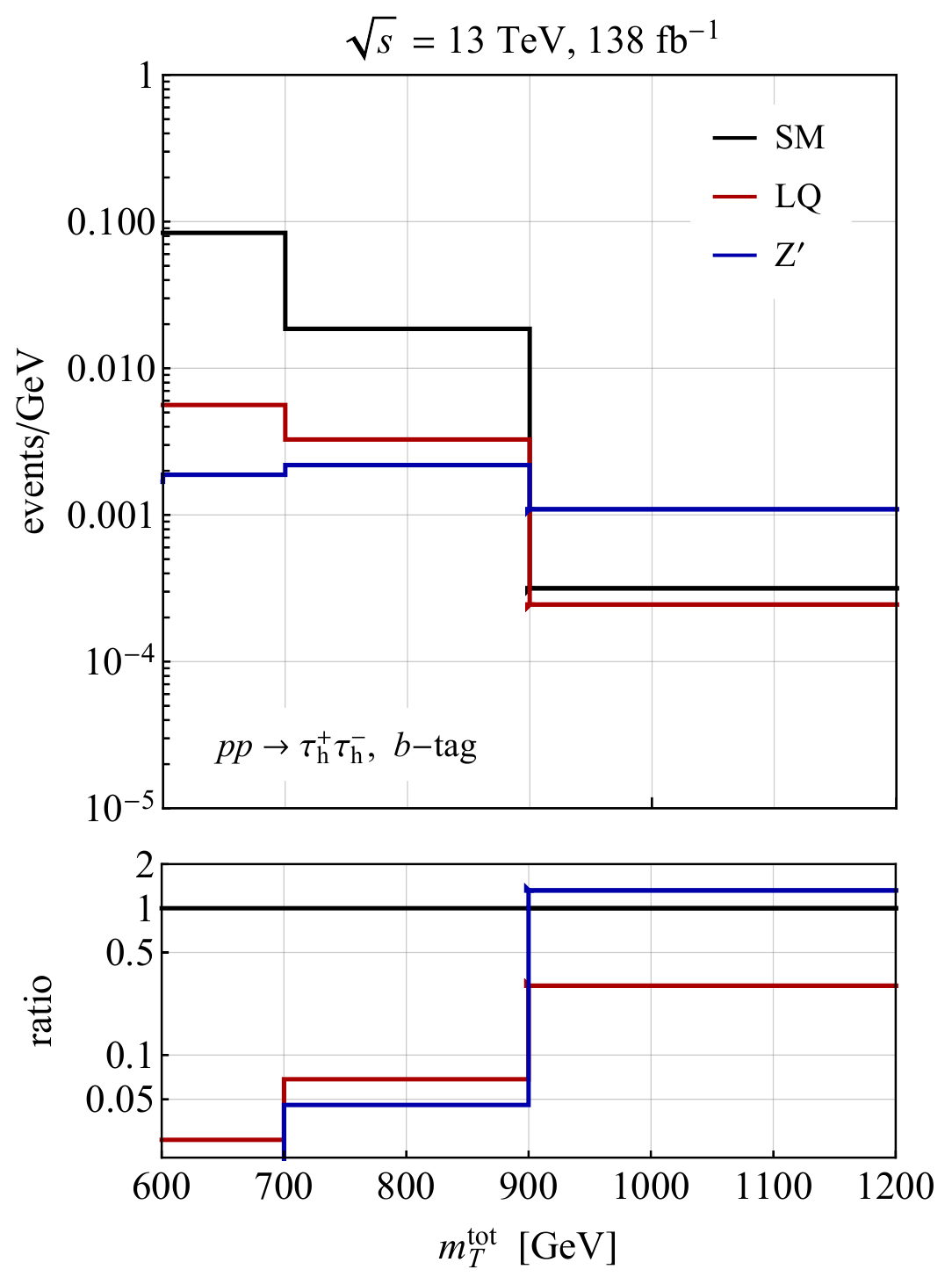}
\end{center}
\vspace{-4mm} 
\caption{\label{fig:phenoZprime} As Figure~\ref{fig:pheno3} but comparing a LQ and a $Z^\prime$ signal hypothesis. The~black curves correspond to the SM expectations of the DY~background provided by~CMS in the publication~\cite{CMS:2022goy}. The red curves  represent the~LQ~NLO predictions assuming $g_4 = 1$,  $M_U = 2 \, {\rm TeV}$ and $M_{G^{\hspace{0.2mm} \prime}} = 2 \, {\rm TeV}$, while the~blue histograms illustrate the LO $Z^\prime$ predictions for  $g_Z^\prime = 1$ and $M_{Z^\prime} = 2 \, {\rm TeV}$. Further details such as the choice of the flavour-dependent $Z^\prime$-boson couplings $\zeta_\psi^{ij}$ can be found in the main~text.}
\end{figure} 

In this appendix we study the possible impact of $Z^\prime$ exchange in DY~ditau production. Following~\cite{Baker:2019sli} we parametrise the interactions between the  colour singlet state~$Z^\prime \sim \left(1,1, 0 \right)$ that appears in the spectrum of the 4321 model after spontaneous symmetry breaking and the SM fermions by 
\begin{equation} \label{eq:LZprime}
{\cal L}_{Z^\prime} \supset  \, \frac{g_{Z^\prime}}{2 \sqrt{6}} \left[\sum_{q=Q,u,d} \zeta_q^{ij} \, \bar{q}^{i} \hspace{0.4mm} \gamma_\mu \hspace{0.4mm} q^j  - 3 \sum_{\ell=L,e} \zeta_\ell^{ij} \, \bar{\ell}^{i} \hspace{0.4mm} \gamma_\mu \hspace{0.4mm} \ell^j  \right] Z^{\prime \mu} \,, 
\end{equation}
where $g_{Z^\prime}$ represents the overall coupling strength of the new neutral gauge boson to SM~matter fields, while~$\zeta^{ij}_{\psi}$ with  $\psi = Q,u,d,L,e$~are $3\times 3$ matrices in flavour space. The observed  semi-leptonic~$B$-decay anomalies can naturally be fulfilled for~$g_{Z^\prime} ={\cal O} (1)$ and $\big |\zeta_\psi^{33} \big |  = {\cal O} (1)$, while the remaining flavour-dependent couplings can be small or vanish identically. 

In Figure~\ref{fig:phenoZprime} we display $m_T^{\rm tot}$ distributions~(\ref{eq:mTtot}) assuming an LQ and a $Z^\prime$ signal hypothesis. For comparison, the SM expectations of the~DY~background taken from~\cite{CMS:2022goy} are also shown as black histograms. Details on the CMS search and our analysis chain can be found at the beginning of~Section~\ref{sec:limits}. The red curves  are the LQ~NLO predictions obtained using our {\tt POWHEG-BOX} code and they employ the parameter choices $g_4 = 1$,  $M_U = 2 \, {\rm TeV}$ and $M_{G^{\hspace{0.2mm} \prime}} = 2 \, {\rm TeV}$. The~$Z^\prime$ predictions have instead been obtained at LO using {\tt MadGraph5\_aMC\@NLO} together with the implementation of~(\ref{eq:LZprime}) provided in the article~\cite{Baker:2019sli}. Our $Z^\prime$-boson event samples correspond to $g_Z^\prime = 1$, $\zeta_\psi^{33} =1$ and $M_{Z^\prime} = 2 \, {\rm TeV}$, while setting all remaining flavour-dependent couplings  $\zeta_\psi^{ij}$ to zero. From both panels one observes that the $m_T^{\rm tot}$ spectra of the $Z^\prime$ signal are on average harder than the distributions resulting from~LQ~exchange. This is expected because the $Z^\prime$ signal arises from $s$-channel exchange, while the LQ contributions are dominantly associated to $t$-channel scattering. It is also evident from the two plots that a simple cut-and-count analysis  based on the observable $m_T^{\rm tot}$ will only have limited power to distinguish between a LQ and a $Z^\prime$ hypothesis. Multivariate discriminants that incorporate the event kinematics of the selected ditau events in both the no $b\hspace{0.4mm}$-tag and the $b\hspace{0.4mm}$-tag category are likely to enhance the sensitivity to different realisations of the 4321 model. A dedicated analysis of this issue is however clearly beyond the scope~of this appendix. 

\end{appendix}


%

\end{document}